\documentclass[
    10pt,
    aps,
    pra,
    preprint,
    superscriptaddress,
    amsmath,
    amssymb
]{revtex4-2}

\usepackage{bm}
\usepackage{graphicx}
\usepackage{physics}
\usepackage{xcolor}
\usepackage{hyperref}
\usepackage{float}

\newcommand{\Trv}{\operatorname{Tr}_{\mathrm{vib}}}

\newcommand{\Imm}{\operatorname{Im}}
\newcommand{\wc}{\omega_{\mathrm c}}

\newcommand{\Hb}{H_{\mathrm B}}
\newcommand{\Hsb}{H_{\mathrm{SB}}}
\newcommand{\Hs}{H_{\mathrm S}}
\newcommand{\Hc}{H_{\mathrm L}}
\newcommand{\Lred}{\mathcal{R}}
\newcommand{\Ldag}{\mathcal{R}^{\dagger}}
\newcommand{\Cgb}{C_{gb}}
\newcommand{\Jopt}{J}
\newcommand{\eps}{\varepsilon}

\begin{document}

\title{Molecular Vibronic Complexity as a Quantum Resource for Optimal Control: Nonadiabatic Mixing Creates Protected Coherence Subspaces in Dissipative Environments}

\author{Elisa Palacino-Gonz\'alez}
\email{elisa.palacino.gonzalez@hhu.de}


\affiliation{
Theoretical and Computational Chemistry Institute,
Heinrich-Heine-Universit\"at D\"usseldorf,
Universit\"atsstrasse 1, 40225 D\"usseldorf, Germany
}


\begin{abstract}

In molecules, electronic coherence typically decays within a few hundred femtoseconds, a timescale usually attributed to the environment. We show that it is set to a large extent by which coherences are excited: the molecule can host coherences that the bath barely sees, and optimal control can find and exploit them. Using gradient-based optimal control (GRAPE) with Redfield dynamics, we maximize the total vibronic coherence between the electronic ground and an optically bright electronic excited state in a three-state model with nonadiabatic features, in two regimes of low and high nonadiabatic mixing and for different system-bath coupling strengths. The Liouvillian spectrum reveals, in both regimes, a well-separated coherence mode with substantial ground-bright character that decays on a picosecond timescale. This mode is protected by mixing with the dark state, which brings the geometry of the optically accessible excited state close to that of the ground state and thus weakens its coupling to the bath. Without prior information, GRAPE finds this mode and keeps the photoexcited wave packet near the Franck-Condon region. At equal fluence, it increases the coherence at 800 fs by more than two orders of magnitude over a Gaussian pulse at the weakest coupling, and by many more at stronger coupling. The Liouvillian predicts the bath dependence of the long-lived coherence to within 1\%, and a Gaussian pulse tuned to the mode frequency comes within 10\% of the best pulse found. Beyond two-level and spin systems, our study puts forward the vibronic structure of molecules as a route to protected coherence subspaces, identifying a dark-state protection mechanism and a design criterion based on molecular parameters. These results lay the groundwork for our goal of using vibronic molecular systems as a resource for long-lived, controllable coherences in molecular quantum information science.

\end{abstract}


\maketitle




\section{Introduction}
\label{sec:introduction}

Electronic coherence in molecules, the phase relationship between electronic states, is written by photoexcitation and rewritten, continuously, by the surrounding bath \cite{schultz_coherence_2024}. Following excitation, coherent superpositions of electronic states evolve together with nuclear wave packets \cite{DomckeCederbaum,WIREPalacino,gatti_novel_2006, Falke2014}, so that electronic and nuclear degrees of freedom often are treated at the same time. Every open quantum system is, by definition, coupled to an environment, which is responsible for the loss of quantum mechanical effects such as decoherence \cite{brif_decoherence_2001,BreuerPetruccione,WeissBook}. In molecules, this can be also interpret as loss of information: the phase imprinted by the photoexcitation in the molecular subspace is lost, typically within a few hundred femtoseconds \cite{Scholes2017,Cao2020,scherer1991fluorescence,hildner2011femtosecond}. In this context, the bath is usually treated as a fixed property that sets the lifetime of the coherence. But is this lifetime really fixed by the bath alone, or does the molecule itself offer coherences that the bath barely sees, which a suitably shaped field could address?
\vspace{0.1cm}

Coherent control, and more specifically optimal control theory \cite{Lecompte_Control, Veintemillas2025, sola_optimal_1998, balint-kurti_optimal_2008, nuernberger_femtosecond_2007,hoff_optimal_2012, PhysRevLett.121.060401,Tiwarireview}, offers a way to approach this question. Foundational theoretical work by Tannor and Brumer established that tailored optical pulses can steer molecular wave-packet dynamics \cite{brumer_coherent_1986, tannor1986coherent,Brumer1986CPL}. Optimal control theory (OCT) subsequently supplied a general mathematical machinery for turning a target objective directly into a control field. Among the different OCT algorithms, Gradient Ascent Pulse Engineering (GRAPE) \cite{Khaneja2005}, Zhu-Rabitz type \cite{Rabitzcontrol} and Krotov \cite{Krotovmethod} methods have been the most used in theoretical molecular physics studies \cite{Rabitz1995,Palao2008,Koch2016,Koch2025}. Most of these studies target populations, such as population transfer or the preparation of target states. Additionally, Koch and coworkers \cite{Palao2008} introduced an indirect approach to protect coherences through state-protection mechanisms, in which the dissipation is represented by population loss from forbidden states rather than by a thermal bath, with gate operations targeting coherences also presented in the context of quantum information processing \cite{Stephane}. This included molecular subspaces based on vibrational states \cite{tesch_quantum_2002,tesch_vibrational_2004,mishima_quantum_2008,berrios_high_2012}. Optimal control has also been shown to exploit, rather than only counteract, dissipation in open quantum systems \cite{Schmidt2011}, and to address the decoherence of molecular vibrational wave packets \cite{brif_decoherence_2001}. The control of decoherence itself has been demonstrated experimentally for vibrational wave packets in diatomic molecules \cite{Branderhorst2008}, laser control of electronic decoherence arising from the loss of nuclear wave-packet overlap has been proposed \cite{Hu2020}, and dark, relaxation-free subspaces have been identified as optimal targets for population transfer \cite{Yuan2012}. However, these studies have focused predominantly on spin-based qubits, atomic models or isolated molecular degrees of freedom, and the direct optimization of electronic coherence in dissipative vibronic molecules has not been explored yet in detail.
\vspace{0.1cm}

In fact, optimizing coherence directly is a more delicate problem than optimizing population. Coherence has both amplitude and phase, and in a vibronic molecule it has nuclear structure as well. On top of that, the interaction with a dissipative environment, which in molecules is rarely negligible, continuously dephases the quantity we aim to control. A control field can mainly act on the coherence through the phase and the shape of the wave packet it creates during the pulse. In a model with displaced potentials the problem becomes even more interesting: the optical excitation initiates nuclear motion on the displaced excited state, and the resulting wave packet sweeps through the phase space while the environment dephases it. The present work builds on our recent studies on coherent control using chirped-pulse excitation \cite{JiaHao,Strich2025}, in which the analysis of nuclear wave-packet Wigner functions showed that vibronic coherence lifetimes can be substantially extended in dissipative vibronic molecules via pulse-chirping. This effect arises from the coherent manipulation of the dissipative and dynamical phases of the vibronic wave packet in the displaced excited state by excitation with a pulse of quadratic temporal phase. In a subsequent study in collaboration with experimental partners, we demonstrated that chirp-dependent double-pump photoluminescence spectra can selectively reveal and manipulate the vibronic dynamics of low-frequency modes in an excitonic conjugated polymer \cite{Xu2025, Malinovskayachirpedcontrol}. Together, these results showed that the temporal structure of the field represents a genuine tool on vibronic coherence. The next step is to ask whether an OCT-optimized field can do better and, more importantly, by what mechanism.
\vspace{0.1cm}

This raises the question addressed in this work: to what extent can an optical excitation extend the lifetime of electronic (and therefore vibronic) coherence beyond the timescale imposed by the intrinsic dissipative dynamics of the molecule? We address it by optimizing the driving field within a dissipative dynamical framework described via Redfield theory. A parallel hypothesis evaluated along this work is whether such control can uncover, and exploit, the role of nominally dark electronic states as a route to control dynamical in vibronic Hilbert spaces.
\vspace{0.1cm}

The GRAPE algorithm \cite{Khaneja2005} is implemented here within the dissipative vibronic Hilbert space of a three-electronic-state nonadiabatic molecular model, comprising a ground state and two electronically coupled excited states. The excited-state potentials are displaced along an explicitly treated vibrational coordinate, which mediates the coupling to a Markovian dissipative environment. The higher-lying excited state is optically accessible from the ground state, whereas the lower-lying one is optically dark. The bath is a linearly coupled harmonic environment with an Ohmic spectral density, treated within Redfield theory. Two models with deliberately different vibronic character are investigated, a low-frequency mode with relatively large Huang-Rhys factors, and a high-frequency mode with moderate coordinate displacements, each at weak, intermediate and strong system-bath coupling. The control objective is introduced as the Frobenius norm of the ground-bright electronic coherence over the vibrational subspace at a fixed target time of 800 fs, chosen well past the few hundred femtoseconds over which the coherence created by a conventional excitation has decayed. Beyond defining a direct coherence target in the vibronic Hilbert space, a further novelty of the present implementation is the use of a Blackman envelope \cite{SolaBlackman}, which keeps the optimized pulses comparable with experimental pulse-shaping techniques.
\vspace{0.1cm}

Three aspects are examined. First, the extent to which optimal control maximizes the long-time electronic coherence relative to a conventional Gaussian excitation at the same fluence, in the two distinct vibronic regimes and for increasing system-bath coupling. Second, the mechanism behind this enhancement, identified through the analysis of the Liouvillian spectrum and the vibronic structure of its slowly decaying modes. Third, whether the physical mechanism found can be used design the excitation \textit{a posteriori}, via GRAPE initialization with a Gaussian pulse tuned to the relevant Liouvillian mode that drives the physical mechanism.
\vspace{0.1cm}

The present work reveals that in both models, the Liouvillian spectrum contains a dominant slowly decaying coherence mode with significant ground-bright (coherence) character through mixing with the dark electronic state. This mixing reduces the displacement between the vibronic states involved, so that the bath has little to dephase. The optimized field can find the mode and populate it, similarly to a Gaussian pulse tuned to its frequency. To our knowledge, this is the first identification of a dark-state-protected electronic coherence mode in a dissipative vibronic model through the Liouvillian spectrum, and the first demonstration that optimal control finds and exploits it. Controlling the coherence then becomes mostly a matter of knowing where this source of long-lived coherence lives in the molecule by analyzing the Liouvillian spectrum, including how its lifetime scales with the coupling to the bath. From this perspective, dark electronic states of a molecule do not need to be seen as a loss channel but a built-in resource, providing coherence components that decay on longer timescales (up to picosecond), in systems whose vibronic structure can be tuned chemically. Optimal control of coherence has so far been addressed mainly for two-level and spin molecular systems. The central message of the present paper is that Vibronic Hilbert spaces thus host coherences that are protected by the electronic and vibrational structure of the molecule. This establishes a starting point for using molecular vibronic systems as carriers of long-lived coherence in quantum information applications\cite{WasielewskiNatRevChem2020,scholes_molecular_2023,scholes_quantum_2025}.


\section{Model}
\label{sec:model}

\subsection{Vibronic Hamiltonian}
\label{subsec:hamiltonian}

A molecular system with three electronic states, $\ket{g}$, $\ket{d}$, and $\ket{b}$, corresponding to the ground, dark excited, and bright excited states, respectively, is considered. The nuclear dynamics is represented by a single explicitly treated harmonic vibrational mode \cite{DomckeCederbaum,KuhnBook2004,PalacinoSMW}. Using dimensionless coordinate and momentum operators $Q$ and $P$, the system Hamiltonian is written as

\begin{equation}
\Hs =
\sum_{\kappa=g,d,b}
\ket{\kappa}\bra{\kappa}
\otimes
\left[
E_\kappa
+
\frac{\hbar\nu}{2}
\left(
P^2+(Q-s_\kappa)^2
\right)
\right]
+
\Delta_{db}
\left(
\ket{d}\bra{b}
+
\ket{b}\bra{d}
\right)
\otimes
\mathbb{I}_{\mathrm{vib}},
\label{eq:system_hamiltonian}
\end{equation}

where $E_k$ and $s_\kappa$ are the electronic energy and dimensionless equilibrium displacement of electronic state $\kappa$, respectively, and $\Delta_{db}$ is the electronic coupling between the dark and bright states.

The ground-state equilibrium is chosen as the origin, $s_g=0$, while the excited-state displacements are $s_d$ and $s_b$. The corresponding potential-energy surfaces are

\begin{equation}
V_\kappa(Q)
=
E_\kappa
+
\frac{\hbar\nu}{2}(Q-s_\kappa)^2.
\label{eq:pes}
\end{equation}

Each electronic state is represented in a basis of the 20 lowest harmonic-oscillator eigenstates of the ground-state mode, in which $Q=(a+a^\dagger)/\sqrt{2}$ and $P=i(a^\dagger-a)/\sqrt{2}$.

The system is initially in thermal equilibrium in the electronic ground state,
\begin{equation}
\rho_0
=
\ket{g}\bra{g}
\otimes
\frac{e^{-\beta H_g}}{\Trv\, e^{-\beta H_g}},
\qquad
H_g = \frac{\hbar\nu}{2}\left(P^2+Q^2\right),
\label{eq:initial_state}
\end{equation}
with $\beta=(k_{\mathrm B}T_{\mathrm{bath}})^{-1}$ and $T_{\mathrm{bath}}=298\ \mathrm{K}$.

\begin{figure}
    \centering
    \includegraphics[width=1.0\textwidth]{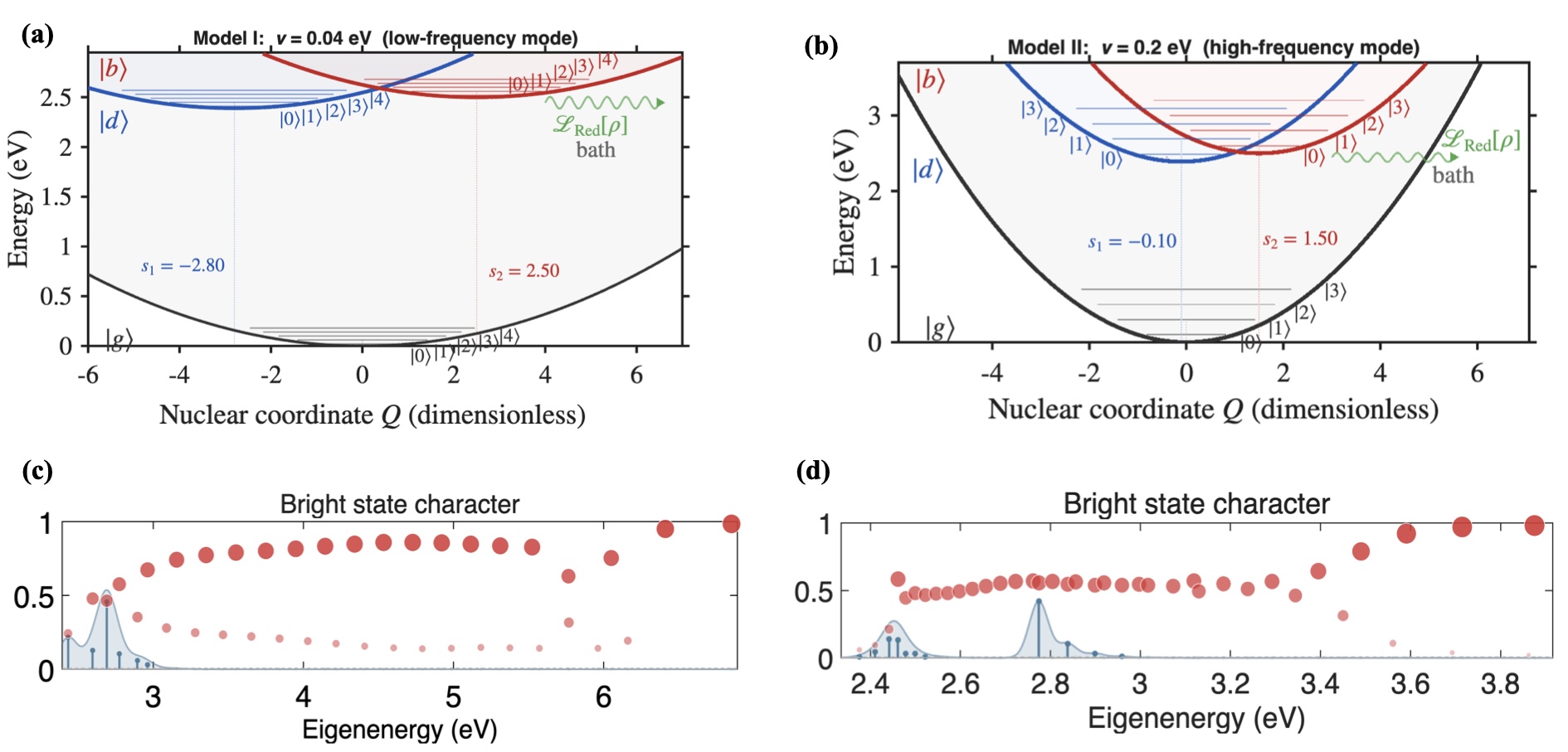}
    \caption{Diabatic potential-energy curves of the ground ($g$), dark ($d$) and bright ($b$) electronic states along the vibrational coordinate $Q$ for (a) Model I ($\hbar\nu=0.04$ eV, $s_d=-2.8$, $s_b=2.5$ - low nonadiabatic mixing) and (b) Model II ($\hbar\nu=0.20$ eV, $s_d=-0.10$, $s_b=1.5$ - high nonadiabatic mixing). The dark and bright states are coupled by $\Delta_{db}=0.15$ eV. Only the bright state is optically accessible from the ground state. Panels (c) and (d) show the energy landscape of eigenstates for both systems, with the vertical axis representing the bright electronic state character defined by the nonadiabatic mixing in the system.}
    \label{fig:vibronic}
\end{figure}

The electronic energies are $E_g=0$, $E_d=2.39\ \mathrm{eV}$, and $E_b=2.50\ \mathrm{eV}$, with a fixed electronic coupling of $\Delta_{db}=0.15\ \mathrm{eV}$. The system parameters chosen are arbitrary and represent typical values of vibronic molecular systems. Two vibronic regimes are investigated (shown in Fig. \ref{fig:vibronic}), chosen deliberately to be as dynamically distinct as possible. The key difference between the two models is the degree of mixing between the nonadiabatically coupled electronic states. This mixing controls how much the dark state contributes to the vibronic eigenstates, and therefore which dissipative pathways govern the dephasing of the electronic coherence.

\paragraph{Model I: low degree of nonadiabatic mixing.}

The first model is characterized by a low vibrational energy, $\hbar\nu=0.04\ \mathrm{eV}$, and large equilibrium displacements, $s_d=-2.8$ and $s_b=2.5$. The low vibrational frequency yields a dense vibronic manifold, and the large displacements drive substantial, slow nuclear motion following excitation, the regime in which one would expect vibronic dephasing to be most punishing.

\paragraph{Model II: high degree of nonadiabatic mixing.}

The second model is characterized by $\hbar\nu=0.20\ \mathrm{eV}$, with $s_d=-0.10$ and $s_b=1.5$. The five-times larger vibrational frequency gives a widely spaced vibronic landscape, and the small dark-state displacement together with the moderate bright-state displacement produces faster, and more compact nuclear dynamics, in contrast to Model I.

The diabatic potential energy curves of the two models considered are shown in Fig. \ref{fig:vibronic}.

\subsection{Light-matter interaction}
\label{subsec:lightmatter}

The molecule is driven by a classical electromagnetic field in the dipole approximation. The total time-dependent system Hamiltonian is given by

\begin{equation}
H(t)=\Hs+\Hc(t),
\qquad
\Hc(t)=-\eps(t)\hat{\mu}.
\label{eq:total_system}
\end{equation}

For the present model, the transition dipole operator couples the ground and bright electronic states,

\begin{equation}
\hat{\mu}
=
\mu_{gb}\,\hat V,
\qquad
\hat V = \hat X + \hat X^\dagger,
\qquad
\hat X = \ket{g}\bra{b}\otimes\mathbb{I}_{\mathrm{vib}},
\label{eq:dipole}
\end{equation}

where $\mu_{gb}$ is the transition dipole moment. The dipole moment and the sign of the interaction are described into a single real control variable, defined as

\begin{equation}
u(t) = -\mu_{gb}\,\eps(t),
\qquad
H(t) = \Hs + u(t)\,\hat V,
\label{eq:control_variable}
\end{equation}

which is the quantity optimized in Sec. \ref{sec:optimal_control}. Throughout, $\mu_{gb}=1$, so that $u(t)$ and $\eps(t)$ are both expressed in energy units (eV).

The initial field is chosen as a Gaussian pulse centred at $t=0$,

\begin{equation}
\eps_0(t)
=
A
\exp
\left(
-\frac{t^2}{2\tau^2}
\right)
\cos
\left(
\frac{\omega_L t}{\hbar}
\right),
\label{eq:initial_pulse}
\end{equation}

with $\tau=25\ \mathrm{fs}$, and $A=1\times10^{-3}\ \mathrm{eV}$. The central carrier frequency of the pulse is set slightly detuned above the minima of the optically-bright excited state, $\omega_L=2.53\ \mathrm{eV}$. This Gaussian pulse serves as the seed from which the reference GRAPE optimization is initialized and represents the transform-limited excitation against which the optimized excitation is compared. In Sec. \ref{subsec:dissipation_results}, the same pulse is later used with its carrier frequency tuned to a specific mode of the Liouvillian spectrum ($\omega_L=2.754$ eV in Model I and $2.317$ eV in Model II), with the pulse fluence unchanged.


\section{Dissipative dynamics}
\label{sec:dissipation}

\subsection{System-bath Hamiltonian}
\label{subsec:systembath}

A reduced density matrix formalism with an explicit system-bath partition is adopted to treat the open quantum dynamics \cite{Palacino2019}. The coupling of the system to a dissipative bath is described by a harmonic bath linearly coupled to the molecular nuclear coordinate. The total Hamiltonian is given by

\begin{equation}
H_{\mathrm{tot}}(t)
=
H(t)
+
\Hb
+
\Hsb,
\label{eq:total_hamiltonian}
\end{equation}

where the bath Hamiltonian is

\begin{equation}
\Hb
=
\sum_{\xi}
\hbar\omega_\xi
b_\xi^\dagger b_\xi
\label{eq:bath_hamiltonian}
\end{equation}

and the system-bath interaction is

\begin{equation}
\Hsb
=
\hat{S}\otimes\hat{B}.
\label{eq:system_bath}
\end{equation}

The bath operator is

\begin{equation}
\hat{B}
=
\sum_\xi
g_\xi
\left(
b_\xi+b_\xi^\dagger
\right),
\label{eq:bath_operator}
\end{equation}

and the system operator is taken to be the dimensionless nuclear coordinate in every electronic state, $\hat{S}=\mathbb{I}_{\mathrm{el}}\otimes Q$, carrying the nuclear structure of the electronic coherence.

The bath is characterized by an Ohmic spectral density with exponential cutoff \cite{WeissBook},

\begin{equation}
J(\omega)
=
\eta\,\omega
\exp
\left(
-\frac{|\omega|}{\wc}
\right),
\label{eq:spectral_density}
\end{equation}

extended to negative frequencies as an odd function, where $\eta$ controls the system-bath coupling strength and $\wc$ is the cutoff. The choice $\wc=\hbar\nu$ is made, matching the bath cutoff to the explicitly treated molecular vibrational energy.

Three dissipation strengths are considered in this work, $\eta=0.03$, $0.10$, and $0.30$, referred to as weak, intermediate, and strong coupling, respectively. Their effect is analysed in Sec. \ref{sec:results}.


\subsection{Redfield dynamics}
\label{subsec:redfield}

Tracing out the bath to second order in $\Hsb$ within the Markovian approximation, the reduced vibronic density operator obeys the Redfield master equation \cite{Redfield,KuhnBook2004,Palacino2019}

\begin{equation}
\frac{d\rho(t)}{dt}
=
-\frac{i}{\hbar}
\left[
H(t),\rho(t)
\right]
+
\Lred[\rho(t)],
\label{eq:redfield_master}
\end{equation}

with $H(t)$ given by Eq. (\ref{eq:control_variable}). The equation is solved in the eigenbasis $\{\ket{m}\}$ of $\Hs$, with eigenenergies $E_m$ and transition energies $\omega_{mn}=E_m-E_n$; the density matrix represented in this basis is denoted $\tilde\rho$. In this basis the Redfield dissipation operator $\Lred$ is written as

\begin{equation}
\Lred[\rho]
=
G_+\rho\, G - G G_+\rho + G\rho\, G_- - \rho\, G_- G
=
-\left[\,G,\; G_+\rho-\rho\, G_+^{\dagger}\,\right],
\label{eq:redfield_operator}
\end{equation}

where $G_{mn}=\bra{m}\hat S\ket{n}$ is the system-bath coupling operator, and the rate-weighted operators are

\begin{equation}
(G_-)_{mn} = G_{mn}\,C_{mn},
\qquad
(G_+)_{mn} = G_{mn}\,C_{nm},
\qquad
G_+^\dagger = G_- .
\label{eq:G_pm}
\end{equation}

The coefficients $C_{mn}$ are the real part of the half-sided Fourier transform of the bath correlation function evaluated at the system transition energies,

\begin{equation}
C_{mn}
=
\mathrm{Re}\int_0^\infty d\tau\, C(\tau)\, e^{i\omega_{mn}\tau/\hbar}
\;\propto\;
J(\omega_{mn})\left[1+n(\omega_{mn})\right],
\label{eq:rates}
\end{equation}

where $n(\omega)=1/(e^{\beta\omega}-1)$ is the Bose-Einstein distribution of the bath at temperature $T_{\mathrm{bath}}$. Explicitly,

\begin{equation}
C_{mn} =
\begin{cases}
J(\omega_{mn})\,\left[1+n(\omega_{mn})\right], & \omega_{mn}>0,\\[4pt]
J(|\omega_{mn}|)\,n(|\omega_{mn}|), & \omega_{mn}<0,\\[4pt]
\eta\, k_{\mathrm B}T_{\mathrm{bath}}, & m=n,
\end{cases}
\label{eq:rates_explicit}
\end{equation}

where the diagonal value is the $\omega\to0$ limit of the off-diagonal expression. The bath correlation function entering Eq. (\ref{eq:rates}) is that of a thermal harmonic bath,

\begin{equation}
C(t)
=
\frac{1}{\pi}
\int_0^\infty
d\omega\,
J(\omega)
\left[
\coth
\left(
\frac{\beta\omega}{2}
\right)
\cos\left(\frac{\omega t}{\hbar}\right)
-
i\sin\left(\frac{\omega t}{\hbar}\right)
\right].
\label{eq:bath_correlation}
\end{equation}

No secular approximation is made and all population-coherence and coherence-coherence transfer terms of the Redfield tensor are retained. The correlation function, together with $\eta$ and $\wc$, fixes the decay rates of the individual vibronic coherences, and hence the timescale over which a coherence prepared by a conventional (uncontrolled) excitation is expected to decay. The vibrational coordinate is retained explicitly in the reduced density matrix, with $\rho(t)$ carrying both electronic populations and coherences and the nuclear structure.


\section{Optimal-control formulation}
\label{sec:optimal_control}

\subsection{Targeting electronic coherences as control objective}
\label{subsec:objective}

The objective functional of our optimal-control framework is defined by the electronic coherence between the ground and bright states at a target time $T$ chosen well after the end of the excitation. The corresponding vibronic coherence operator is defined as

\begin{equation}
\hat{\rho}_{gb}^{\mathrm{vib}}(t)
\equiv
\bra{g}\rho(t)\ket{b},
\label{eq:vibronic_coherence}
\end{equation}

which acts in the full vibrational Hilbert space.

The reduced electronic coherence is obtained by tracing over the nuclear degrees of freedom,

\begin{equation}
\rho_{gb}(t)
=
\Trv
\left[
\hat{\rho}_{gb}^{\mathrm{vib}}(t)
\right],
\label{eq:electronic_coherence}
\end{equation}

and its magnitude is

\begin{equation}
\Cgb(t)
=
\left|
\rho_{gb}(t)
\right|.
\label{eq:coherence_measure}
\end{equation}

As target objective, the Hilbert-Schmidt norm (equivalently, the Frobenius norm in the diabatic vibronic basis) of the vibronic coherence is considered,

\begin{equation}
\Cgb^{\mathrm{vib}}(t)
=
\left[
\Trv
\left(
\hat{\rho}_{gb}^{\mathrm{vib}\dagger}(t)
\hat{\rho}_{gb}^{\mathrm{vib}}(t)
\right)
\right]^{1/2}.
\label{eq:vibronic_norm}
\end{equation}

All results in Sec. \ref{sec:results} are given in terms of $\Cgb^{\mathrm{vib}}$. All quantities below are expressed in terms of the control variable $u(t)=-\mu_{gb}\eps(t)$ and the coupling operator $\hat V$ defined in Eq. (\ref{eq:control_variable}).

The control field $u(t)$ is optimized to maximize the ground-bright electronic coherence at a prescribed target time $T$, chosen to lie well past the timescale over which the bath would ordinarily be expected to have erased it. The objective functional is
\begin{equation}
\Jopt[u]
=
\Cgb^{\mathrm{vib}}(T)
-
\alpha\,
\mathcal{F}[u],
\label{eq:objective}
\end{equation}
where $\alpha$ is a positive fluence-penalty parameter and
\begin{equation}
\mathcal{F}[u]
=
\int_{t_0}^{t_N}
dt\,
u^2(t)
\label{eq:fluence}
\end{equation}
is the integrated pulse fluence over the full simulation window $[t_0,t_N]$. The first term in Eq. \ref{eq:objective} rewards a large ground-bright vibronic coherence norm at the target time; the second term penalizes the pulse fluence and enters the choice of the adaptive step size. The comparison between optimized and reference pulses is in any case made at exactly equal fluence (Sec. \ref{subsec:grape}).

The target time is $T=800\ \mathrm{fs}$, measured from the centre of the initial pulse at $t=0$. The simulation window extends from $t_0=-400\ \mathrm{fs}$ to $t_N=1600\ \mathrm{fs}$.

In addition to the optimized objective, the performance of the control field is quantified using the enhancement factor

\begin{equation}
\mathcal{G}(T)
=
\frac{
\Cgb^{\mathrm{vib,opt}}(T)
}{
\Cgb^{\mathrm{vib,G}}(T)
},
\label{eq:enhancement}
\end{equation}

where the superscripts ``opt'' and ``G'' denote the optimized and Gaussian reference fields, respectively, compared at equal fluence. $\mathcal{G}(T)$ measures how much coherence a shaped field can retain at a fixed time and energy budget, compared with a conventional excitation of the same molecule.


\subsection{GRAPE optimization}
\label{subsec:grape}

The GRAPE algorithm is implemented within an OCT framework \cite{Khaneja2005}. The simulation window is divided into $N$ temporal slices, with $t_k=t_0+k\Delta t$, $\Delta t = 0.1\ \mathrm{fs}$, and the control is taken to be constant within each time slice:

\begin{equation}
u(t)=u_k,
\qquad
t_k\leq t<t_{k+1}.
\label{eq:piecewise}
\end{equation}

For a given value of $u_k$, the Liouvillian governing the density matrix is

\begin{equation}
\mathcal{L}_k\rho
=
-\frac{i}{\hbar}
\left[
\Hs + u_k\hat V,
\rho
\right]
+
\Lred[\rho],
\label{eq:liouvillian}
\end{equation}

where $\Lred$ is the Redfield dissipation superoperator of Eq. (\ref{eq:redfield_operator}).

The state is propagated from the thermal initial state $\rho_0$ of Eq. (\ref{eq:initial_state}) according to
\begin{equation}
\frac{d\rho}{dt} = \mathcal{L}_k\rho,
\qquad t_k\leq t<t_{k+1},
\label{eq:propagator}
\end{equation}
using a fourth-order Runge-Kutta integrator in the eigenbasis of $\Hs$, with the dissipative superoperator $\Lred$ embedded directly inside every propagation step.

To evaluate the gradient of the objective with respect to the control field, an adjoint operator $\Lambda(t)$ is propagated backward from the target time $T$. Its equation of motion is
\begin{equation}
\frac{d\Lambda(t)}{dt}
=
\frac{i}{\hbar}\left[H(t),\Lambda(t)\right]
-\Ldag[\Lambda(t)]
-\Sigma(t),
\label{eq:adjoint}
\end{equation}
where $\Ldag$ is the adjoint of the Redfield dissipator in
Eq. (\ref{eq:redfield_operator}) and $\Sigma(t)$ is a source term acting on
the ground-bright coherence block.

$\Lambda(T)$ is transformed to the eigenbasis of $\Hs$ before the backward propagation, performed with the same fourth-order Runge-Kutta scheme.

The gradient of the coherence term on each time slice is
\begin{equation}
g_k
=
\frac{1}{\hbar}\,\Imm\,\Tr\!\left[\hat V\tilde\rho(t_k)\Lambda(t_k)\right],
\label{eq:gradient}
\end{equation}
which follows from the linear dependence of $H(t)$ on $u$.

Two optimization protocols are used, which differ in how the update, the fluence and the retained pulse are handled. In the reference optimization, the control field is updated iteratively according to

\begin{equation}
u_k^{(n+1)}
=
w_k
\left(
u_k^{(n)}
+
\gamma_n\,
g_k^{(n)}
\right),
\label{eq:update}
\end{equation}

where we have implemented $w_k$, a Blackman window of half-width $75\ \mathrm{fs}$ centred at $t=0$, which confines the control to a finite temporal support. $\gamma_n$ is an adaptive step: $\gamma_{n+1}=0.5\,\gamma_n$ if $\Jopt$ decreases between consecutive iterations and $\gamma_{n+1}=1.05\,\gamma_n$ otherwise, with $\gamma_0=0.01$.
\vspace{0.1cm}

Each GRAPE iteration therefore consists of forward propagation of the system density matrix under the full dissipative Liouvillian, backward propagation of the adjoint state, evaluation of the gradient, and update of the control field.

\paragraph{Reference optimization.}
The optimization is initialized from the Gaussian pulse of Eq. \ref{eq:initial_pulse}) at $\omega_L=2.53$ eV, which also defines the reference fluence. It is run for 40 iterations with $\alpha=10$, by which point the objective functional $J$ has converged, and the last iterate is retained. In this optimization the update additionally contains the term $2\alpha\gamma_n\Delta t\,u_k$, which changes the amplitude by about 2\% per iteration. The fluence is therefore not fixed during the optimization,but only afterwards by rescaling the final field to the fluence of the Gaussian reference,

\begin{equation}
u^{\mathrm{opt}}_k
\;\rightarrow\;
u^{\mathrm{opt}}_k
\sqrt{
\frac{\mathcal{F}[u^{\mathrm{G}}]}{\mathcal{F}[u^{\mathrm{opt}}]}
},
\label{eq:renormalization}
\end{equation}

and propagated once more, so that the enhancement factor $\mathcal{G}(T)$ in Eq. (\ref{eq:enhancement}) compares both fields at exactly the same fluence.

\paragraph{Liouvillian-seeded optimization.}
Once the protected mode has been identified (Sec. \ref{subsec:liouvillian_spectrum}), the optimization can be reformulated more strictly: starting from the pulse tuned to this mode, at fixed fluence throughout, and keeping the best iterate. In this second set of optimizations, GRAPE is initialized with the Gaussian pulse tuned to the protected mode ($\omega_L = 2.754$ eV in Model I and $2.317$ eV in Model II), with the same envelope and fluence as the reference pulse. In contrast to the reference optimization, no fluence penalty is used ($\alpha=0$), the Blackman window is applied to the gradient rather than to the full pulse, and the field is rescaled to the fluence of the seed after every update,

\begin{equation}
u_k^{(n+1)} =\left(u_k^{(n)} + \gamma_n\,w_k\,g_k^{(n)} \right) \sqrt{ \frac{\mathcal{F}[u^{\mathrm{seed}}]}{\mathcal{F}[u^{(n)}+\gamma_n w g^{(n)}]}},
\label{eq:update_seeded}
\end{equation}

so that every iterate has exactly the reference fluence. The adaptive step $\gamma_n$ follows the same rule as above, but is controlled by $\Cgb^{\mathrm{vib}}(T)$ instead of $\Jopt$. The coherence at $T$ is evaluated for every iterate, and the best pulse over 40 iterations is retained. In all cases, the coherence at $T$ converged within the iteration 15.
With this choice, the seeded optimization cannot perform worse than its initial pulse, and all comparisons are made at identical fluence throughout.


\section{Results}
\label{sec:results}

\subsection{Optimal control enhances long-time electronic coherence}
\label{subsec:coherence_enhancement}

\begin{figure}[H]
    \centering
    \includegraphics[width=1.0\textwidth]{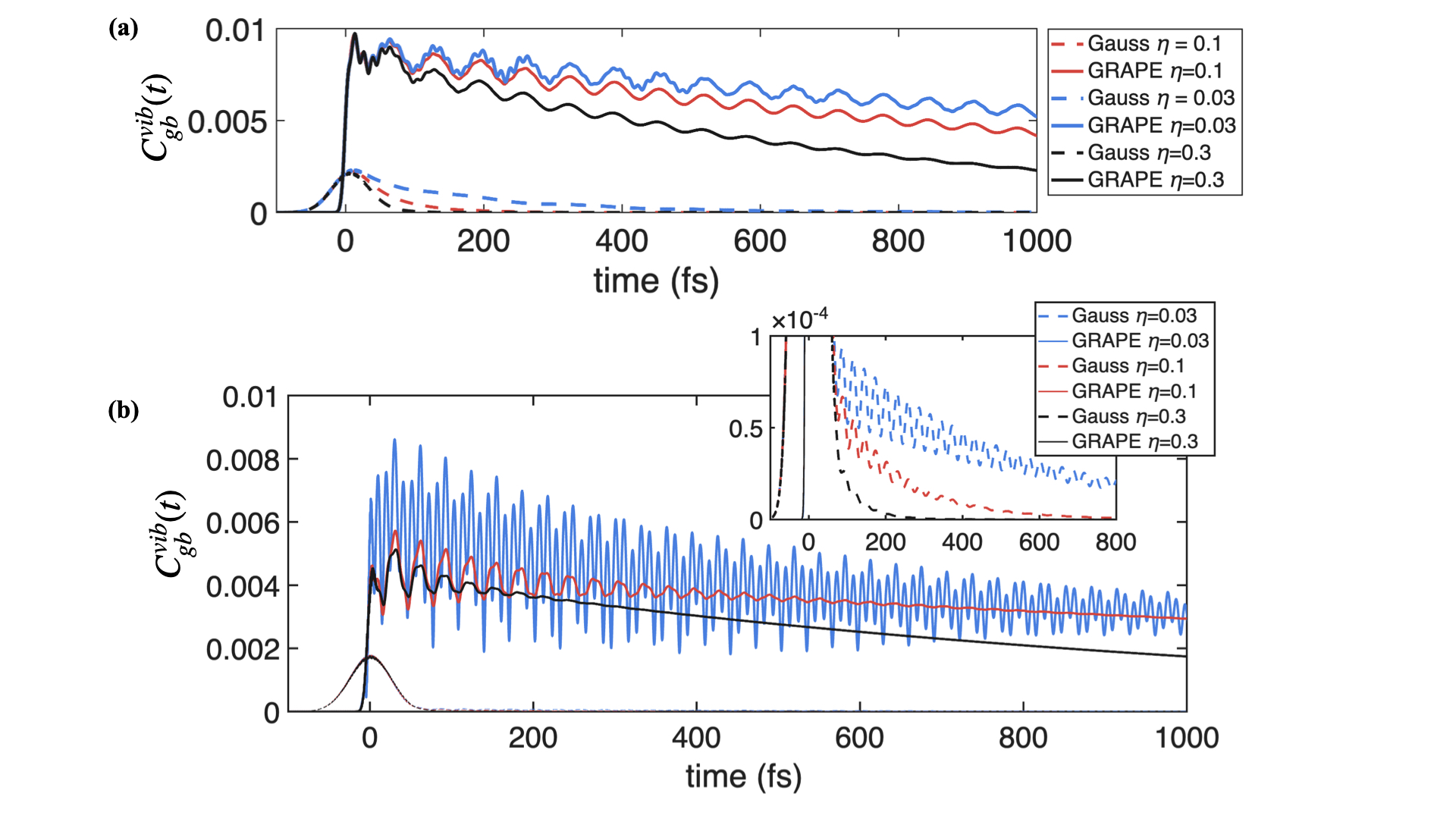}
    \caption{Ground-bright vibronic coherence $\Cgb^{\mathrm{vib}}(t)$ under the Gaussian reference pulse and the GRAPE-optimized pulse, at equal fluence, for $\eta=0.03$, $0.10$ and $0.30$. (a) Model I. (b) Model II. The target time is set to $T=800$ fs.}
    \label{fig:coherences}
\end{figure}

Once the optimal pulse is obtained from the GRAPE optimization, Eq. (\ref{eq:redfield_master}) is propagated under both the Gaussian reference pulse and the control pulse for the two vibronic models. The resulting ground-bright coherence dynamics are compared in Fig. \ref{fig:coherences} for the three system-bath coupling strengths.

For Model I (upper panel)), under the Gaussian reference pulse, the electronic coherence rises briefly after photoexcitation and then decays, accompanied by typical oscillatory nuclear dynamics, with a decay time below 400 fs depending on the bath strength, well inside the $T=800$ fs target window. The control field changes this picture. Panel (a) shows that the control solution is a more compact excitation in time, initially preparing a substantially larger electronic coherence, again followed by pronounced vibronic oscillations. Short-period modulations, arising from the different vibrational coherences prepared in the bright state, are superimposed on the decay and are damped more rapidly with increasing bath coupling. As later discussed in Sec. \ref{subsec:pulses}, the GRAPE optimal field has a broader spectrum and a carrier frequency shifted with respect to the Gaussian reference. These effects translate into a substantially larger electronic coherence that survives beyond the target time. For the weakest bath strength included in this work ($\eta = 0.03$), the electronic coherence survives beyond $1\ \mathrm{ps}$.
\vspace{0.1cm}

The enhancement factor $\mathcal{G}(T)$ grows with the system-bath coupling, from about 170 at $\eta=0.03$ to about $10^5$ at $\eta=0.10$ and $10^9$ at $\eta=0.30$ in Model I. This growth does not mean that the control performs better at strong coupling. The optimized coherence also decreases with $\eta$, but only by a factor of about two, whereas the reference coherence has essentially vanished at $T$. The large values of $\mathcal{G}$ therefore reflect the collapse of the reference coherence picture rather than an improvement of the control pulse. The values for both models are given in Table \ref{tab:protocols} of Subsection \ref{subsec:dissipation_results}.
\vspace{0.1cm}

Model II (panel (b)) shows qualitatively different dynamics. The larger vibrational frequency produces faster oscillatory motion of the vibronic coherences. This is clearly observed for the Gaussian excitation in the close-up at the top right corner, where $\Cgb^{\mathrm{vib}}(t)$ exhibits pronounced oscillations at the vibrational frequency. The optimized GRAPE solution modifies both the amplitude and the phase of these oscillations, leading to a significantly larger coherence at the target time. A similar trend is observed for all bath strengths.
\vspace{0.1cm}

Since the pulses are compared at equal fluence, the advantage does not come from a stronger excitation, but from the spectral and temporal structure of the control field. As described later in Subsecs. \ref{subsec:liouvillian_spectrum} and \ref{subsec:dissipation_results}, the properties of the control pulse determine which modes of the molecular Liouvillian spectrum are selectively populated , as well as the position of the photoinduced wave packet (Subsec. \ref{subsec:wavepacket}).

\subsection{The optimized field adapts to the vibronic dynamics}
\label{subsec:pulses}

\begin{figure}[H]
    \centering
    \includegraphics[width=0.7\textwidth]{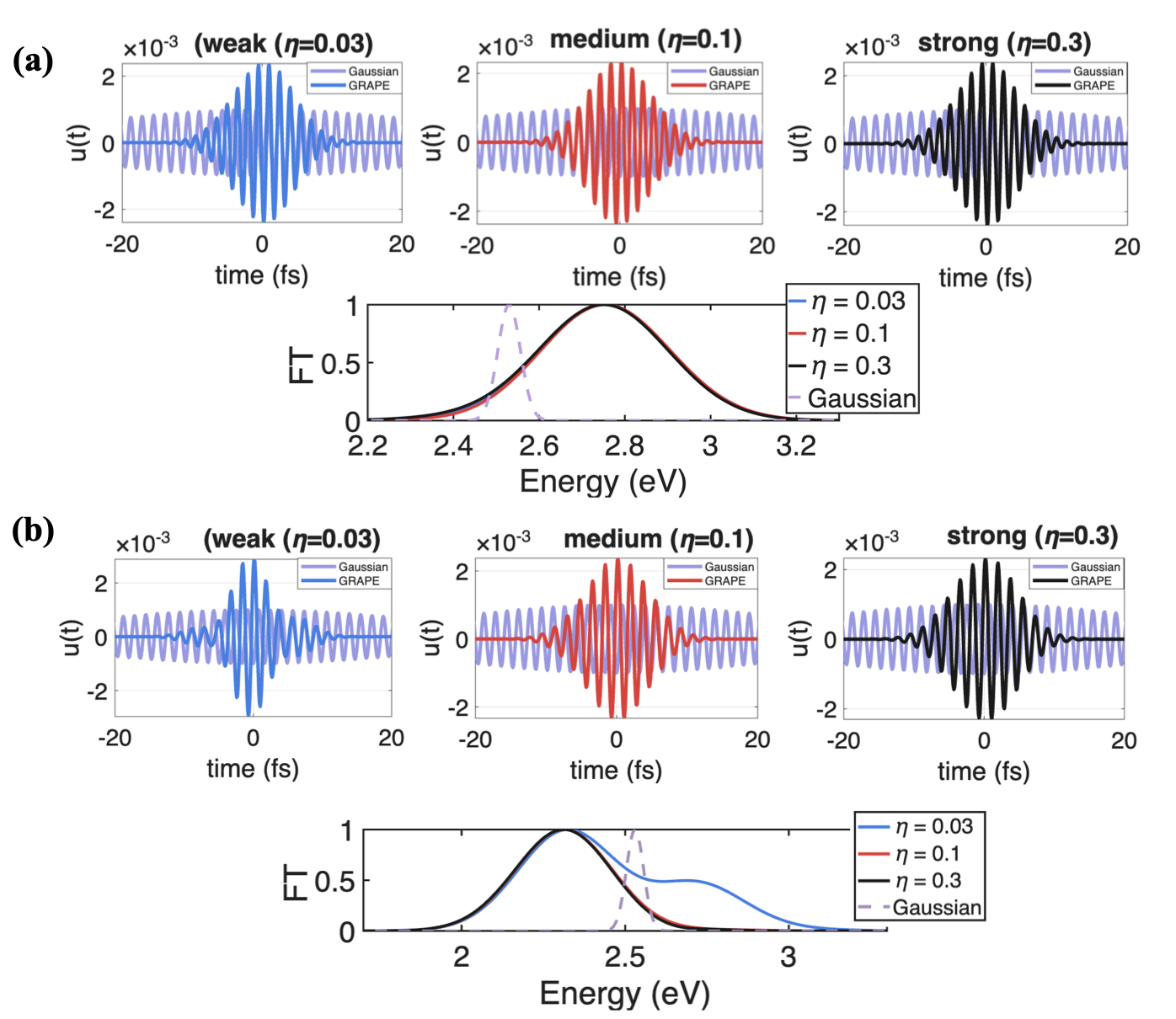}
    \caption{Reference Gaussian (purple) and GRAPE-optimized (black, red, blue) pulses for Models I (a) and II (b), with their Fourier transforms. The optimized fields develop temporal and spectral substructure adapted to the vibronic dynamics and dissipation strength.}
    \label{fig:pulses}
\end{figure}

The control pulses and their Fourier transform are compared with those of the reference Gaussian excitation in Fig. \ref{fig:pulses}. The upper panels (a) show the reference Gaussian pulse (purple), together with the optimized fields and their Fourier transforms, for Model I at the three dissipation strengths. In the time domain, GRAPE consistently leads to a shorter excitation (broader frequency content) with a carrier frequency shifted toward the blue relative to the reference Gaussian. This qualitative behaviour is largely preserved across the different bath strengths. The larger spectral broadening of the control pulse indicates that the photoexcitation prepares the nuclear wave packet over a broader energy window, centered at higher excitation energies. For Model I, the control strategy seems therefore to be relatively insensitive to the magnitude of the dissipation, and is determined mainly by the vibronic landscape of the molecule.
\vspace{0.1cm}

A different picture is observed for Model II (panels (b)), where substantially slower temporal modulations are observed in the control fields. Moreover, here, the pulse structure depends more strongly on the dissipative environment. In the weak-bath regime (blue), the control field develops a shoulder in its Fourier transform. The resulting spectrum contains two prominent frequency regions, below and above the central frequency of the reference Gaussian. The higher-frequency component, centred around 2.7 eV, overlaps with higher vibrational levels of the bright excited-state potential. The red-shifted component, around 2.4 eV, lies below the bright-state origin, in the region where the dark state mixes with the bright state. This additional spectral component progressively disappears as the dissipation strength increases.
\vspace{0.1cm}

The two models thus lead to opposite control strategies: in Model I the optimized field is shifted toward higher energies, whereas in Model II it places substantial spectral weight below the minimum of the bright excited-state potential. Both shifts point toward the frequency of the protected coherence mode later described in Sec. \ref{subsec:liouvillian_spectrum}, which lies at $2.754$ eV in Model I and at $2.317$ eV in Model II, above and below the reference carrier at $2.53$ eV, respectively.

\subsection{The optimized field keeps the excited-state wave packet in the Franck-Condon region}
\label{subsec:wavepacket}

\begin{figure}[H]
    \centering
    \includegraphics[width=0.9\textwidth]{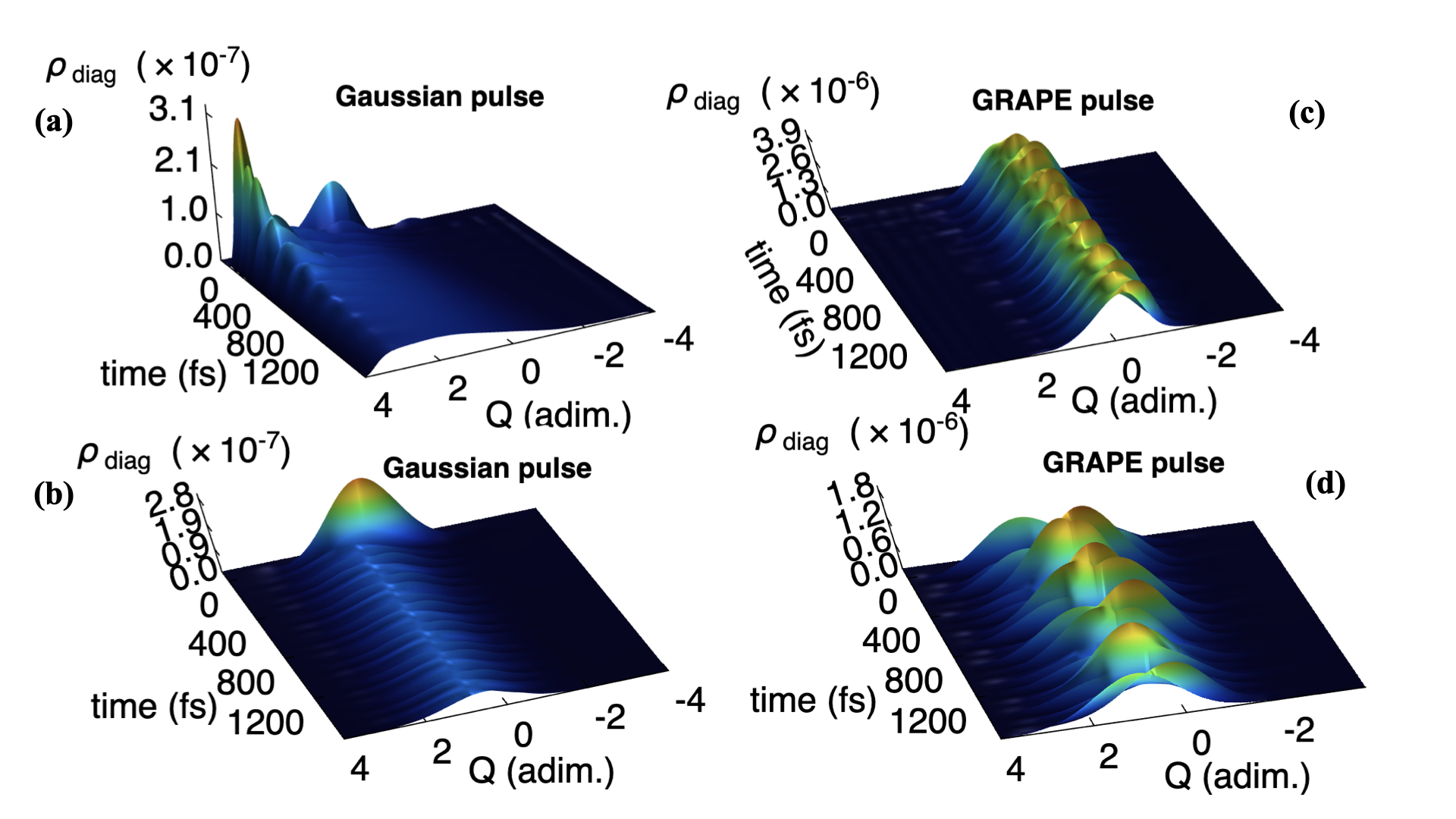}
    \caption{Vibrational density of the bright state, $\rho_{bb}(Q,t)$, after excitation with the Gaussian reference pulse (a, b) and the GRAPE-optimized pulse (c, d), at equal fluence and $\eta=0.03$. (a, c) Model I. (b, d) Model II. Under the reference pulse the wave packet moves to the minimum of the bright-state potential ($Q=s_b$); under the optimized pulse it remains near the Franck-Condon region ($Q\approx0$). Note the different vertical scales.}
    \label{fig:wavepacket}
\end{figure}

Further insight into the effect of the optimized field is obtained from the vibrational density of the bright state,
\begin{equation}
\rho_{bb}(Q,t)=\sum_{v,v'}\chi_v(Q)\,\rho_{bv,bv'}(t)\,\chi_{v'}(Q),
\label{eq:rho_bb_Q}
\end{equation}
where $\chi_v(Q)$ are the harmonic-oscillator eigenfunctions and $\rho_{bv,bv'}(t)=\bra{b,v}\rho(t)\ket{b,v'}$, shown in Fig. \ref{fig:wavepacket} for the Gaussian reference pulse and the GRAPE-optimized pulse at $\eta=0.03$. Under the reference pulse, the wave packet created in the Franck-Condon region moves towards the minimum of the bright-state potential, at $Q=s_b$ ($2.5$ in Model I and $1.5$ in Model II), and the bright-state density has essentially disappeared after about 400 fs (panels a and b). The control pulse produces a qualitatively different picture. In both models the wave packet remains close to the Franck-Condon region, near the ground-state equilibrium geometry at $Q=0$, and the bright-state density persists over the whole simulation window, beyond 1200 fs, with regular vibrational oscillations (panels c and d).
\vspace{0.1cm}

Since the bath couples to the nuclear coordinate, the ground-bright vibronic coherence is dephased to the extent that the nuclear wave packets associated  through both reference excitation, minimum at $Q=s_b$, far from the ground-state geometry, and as a result the coherence is rapidly lost. Interestingly, the control solution behaves in a qualitatively different way. Fig. \ref{fig:wavepacket} reveals that, under the control field, the photoinduced bright-state wavepacket  remains localized near $Q=0$, where the two components of the coherence  nearly indistinguishable to the bath. The larger initial coherence prepared by the optimized field partly reflects its stronger spectral overlap with the relevant vibronic transitions. Notably, however, the decisive difference is not the magnitude of the coherence, but where the wave packet  located and, consequently, how long the coherence survives. This mechanism is made precise in Subsec. \ref{subsec:liouvillian_theory} in terms of the Liouvillian spectrum.

\subsection{The Liouvillian spectrum as a steering picture for control}
\label{subsec:liouvillian_theory}

The dissipative landscape on which the control acts is characterized by the field-free generator $\mathcal{L}_0$, obtained from Eq. (\ref{eq:liouvillian}). Since the Redfield tensor is constructed in the eigenbasis of $\Hs$ and does not depend on the field, the generator of each time slice is given by

\begin{equation}
\mathcal{L}_k=\mathcal{L}_0+\eps_k\,\mathcal{L}_\mu,
\qquad
\mathcal{L}_\mu\rho=\frac{i}{\hbar}\left[\hat{\mu},\rho\right].
\label{eq:liouvillian_split}
\end{equation}

The slice generators $\mathcal{L}_k$ also have well-defined spectra. However, their eigenbases change from slice to slice and, since the slice duration is short compared with the dissipative timescales, they do not provide a reference frame for the dynamics. The spectrum of $\mathcal{L}_0$ is therefore used throughout.

Because $\mathcal{L}_0$ is not Hermitian, right and left eigenoperators are introduced \cite{Minganti2018},

\begin{equation}
\mathcal{L}_0\,\sigma_j=\lambda_j\,\sigma_j,
\qquad
\mathcal{L}_0^{\dagger}\,\tilde{\sigma}_j=\lambda_j^{*}\,\tilde{\sigma}_j,
\qquad
\Tr\big(\tilde{\sigma}_j^{\dagger}\sigma_k\big)=\delta_{jk},
\label{eq:liouvillian_eigen}
\end{equation}

assuming $\mathcal{L}_0$ is diagonalizable. The eigenvalues are written as

\begin{equation}
\lambda_j=-\kappa_j-i\Omega_j,
\label{eq:liouvillian_eigenvalue}
\end{equation}

where $\kappa_j$ is the decay rate and $\Omega_j$ the oscillation frequency of mode $j$. The spectrum of $\mathcal{L}_0$ inherits a block structure from the model. Since neither $\Hs$ nor the system-bath operator $\hat{S}=Q$ couples the ground electronic state to the excited manifold $\{\ket{d},\ket{b}\}$, $\mathcal{L}_0$ maps each electronic block of the density operator onto itself: the ground state ($gg$), the excited manifold ($ee$), and the ground-excited coherences ($ge$ and $eg$). Every eigenoperator $\sigma_j$ can therefore be chosen within a single block, and only the dipole term $\mathcal{L}_\mu$ connects different blocks. Since the populations of the ground and excited manifolds are separately conserved by $\mathcal{L}_0$, the eigenvalue $\lambda=0$ is twofold degenerate, with stationary states in the $gg$ and $ee$ blocks. All remaining eigenoperators are traceless.

Expanding the density operator in the eigenoperators of $\mathcal{L}_0$,

\begin{equation}
\rho(t)=\sum_j c_j(t)\,\sigma_j,
\qquad
c_j(t)=\Tr\big(\tilde{\sigma}_j^{\dagger}\rho(t)\big),
\label{eq:mode_expansion}
\end{equation}
the equation of motion becomes
\begin{equation}
\dot{c}_j(t)=\lambda_j\,c_j(t)+\eps(t)\sum_k M_{jk}\,c_k(t),
\qquad
M_{jk}=\Tr\big(\tilde{\sigma}_j^{\dagger}\mathcal{L}_\mu\sigma_k\big).
\label{eq:mode_eom}
\end{equation}

Free propagation under $\mathcal{L}_0$ is block-diagonal: populations and vibrational coherences remain in the $gg$ and $ee$ blocks, and optical coherences remain in the $ge$ and $eg$ blocks. Each field interaction, through $M_{jk}$, changes one electronic index of the density operator, as in a double-sided Feynman diagram \cite{Mukamelbook}. Between interactions, every mode evolves independently, oscillating at $\Omega_j$ and decaying at $\kappa_j$.

The control target, Eq. (\ref{eq:vibronic_norm}), is expressed in the same basis. Only optical-coherence modes of the $ge$ block contribute to the vibronic coherence operator,

\begin{equation}
\hat{\rho}_{gb}^{\mathrm{vib}}(t)=\sum_{j\in ge}c_j(t)\,s_j,
\qquad
s_j=\bra{g}\sigma_j\ket{b},
\label{eq:vib_coherence_modes}
\end{equation}

and the objective becomes a quadratic form in the mode amplitudes,

\begin{equation}
\big[\Cgb^{\mathrm{vib}}(t)\big]^2
=\sum_{j,k\in ge}c_j^{*}(t)\,c_k(t)\,G_{jk},
\qquad
G_{jk}=\Trv\big(s_j^{\dagger}s_k\big).
\label{eq:objective_modes}
\end{equation}
\vspace{0.1cm}

Populations and vibrational coherences in the $gg$ and $ee$ blocks therefore do not contribute directly to the objective; relaxation processes enter only through the decay rates $\kappa_j$ of the $ge$ modes. Since the eigenoperators of the non-normal generator $\mathcal{L}_0$ are not mutually orthogonal, $G_{jk}$ is not diagonal. Its off-diagonal elements describe interference between optical coherences oscillating at different frequencies, beating at $\Omega_j-\Omega_k$. Over any interval $[t_p,T]$ in which the field vanishes, $c_j(T)=c_j(t_p)\,e^{\lambda_j(T-t_p)}$, so that at long times the objective is dominated by the slowest-decaying optical-coherence modes with appreciable $\|s_j\|$. Because $\Delta_{db}$ mixes the dark and bright states in $\Hs$, these modes carry character from both, which shapes the dissipative landscape experienced by the ground-bright coherence.

The objective $\Cgb^{\mathrm{vib}}(T)$, however, differs from the optical polarization in one important respect. Under the coherent part of $\mathcal{L}_0$, the optical-coherence block evolves as

\begin{equation}
\rho_{ge}(t)=e^{-iH_g t/\hbar}\,\rho_{ge}(0)\,e^{iH_e t/\hbar},
\label{eq:ge_free}
\end{equation}

where $H_g$ and $H_e$ are the restrictions of $\Hs$ to the ground and excited ($d,b$) manifolds. This evolution conserves the Hilbert-Schmidt norm of the full $ge$ block, $\|\rho_{gd}\|^2+\|\rho_{gb}\|^2$. This is not the case for the traced coherence $\Cgb$ of Eq. (\ref{eq:coherence_measure}), which is proportional to the polarization and additionally depends on the overlap between the nuclear wave packets in the ground and bright states. For displaced potentials, this overlap decays and recurs following the nuclear motion, in analogy to the Franck-Condon picture of vibronic absorption. The objective considered here is independent of this overlap. In the absence of the field, $\Cgb^{\mathrm{vib}}$ is only modified by the Redfield dissipator and by the coherent exchange with the ground-dark coherence induced by the electronic coupling $\Delta_{db}$. The optimization therefore acts against bath-induced decoherence and against the redistribution of the coherence into the ground-dark block, rather than against the decay of the wave-packet overlap.
\vspace{0.1cm}

The control mechanism can be further understood at first order in the field. For an initial state $\rho_0$ that is stationary in the ground manifold, Eq. (\ref{eq:mode_eom}) gives

\begin{equation}
\hat{\rho}_{gb}^{\mathrm{vib}}(T)
=\int_{t_0}^T dt\,\eps(t)\,R(T-t)+\mathcal{O}(\eps^3),
\qquad
\hat{R}^{(1)}(\tau)=\sum_{j\in ge}m_j\,e^{\lambda_j\tau}\,s_j,
\label{eq:first_order}
\end{equation}

with
$m_j=\Tr\big(\tilde{\sigma}_j^{\dagger}\mathcal{L}_\mu\rho_0\big)$. $\hat{R}^{(1)}(\tau)$ describes the linear response of the ground-bright coherence to the field. Each $ge$ mode contributes with its frequency $\Omega_j$, its decay rate $\kappa_j$, and a weight $|m_j|\,\|s_j\|$. Modes with $\Trv\,s_j\approx0$ do not contribute to the traced coherence $\Cgb$, but they do contribute to $\Cgb^{\mathrm{vib}}$. At this order, the control problem reduces to the selective excitation of $ge$ modes: at fixed fluence, the control field concentrates its spectral weight on slowly decaying modes with large $|m_j|\,\|s_j\|$, with relative phases such that they interfere between $ge$ modes after the excitation requires coherence-population-coherence pathways and appears only at third order. These contributions are included in the GRAPE optimization, but for the fluences used here the coherence at $T$ is linear in the field amplitude to a very good approximation (Sec. \ref{subsec:dissipation_results}).
\vspace{0.1cm}

This mechanism is closely related to decoherence-free  \cite{Zanardi1997,Lidar1998}. Since the bath couples through $\hat S=\mathbb{I}_{\mathrm{el}}\otimes Q$, identically in all electronic states, it can only distinguish the ground- and bright-state components of the coherence through their nuclear wave packets. Coherences whose two nuclear components are similar along $Q$ are therefore only weakly  Unlike in an exact decoherence-free subspace, however, the protection is  approximate, since the corresponding decay rates $\kappa_j$ are small but finite.

\subsection{A dark-state-protected slow coherence mode}
\label{subsec:liouvillian_spectrum}

\begin{figure}[H]
    \centering
    \includegraphics[width=1.0\textwidth]{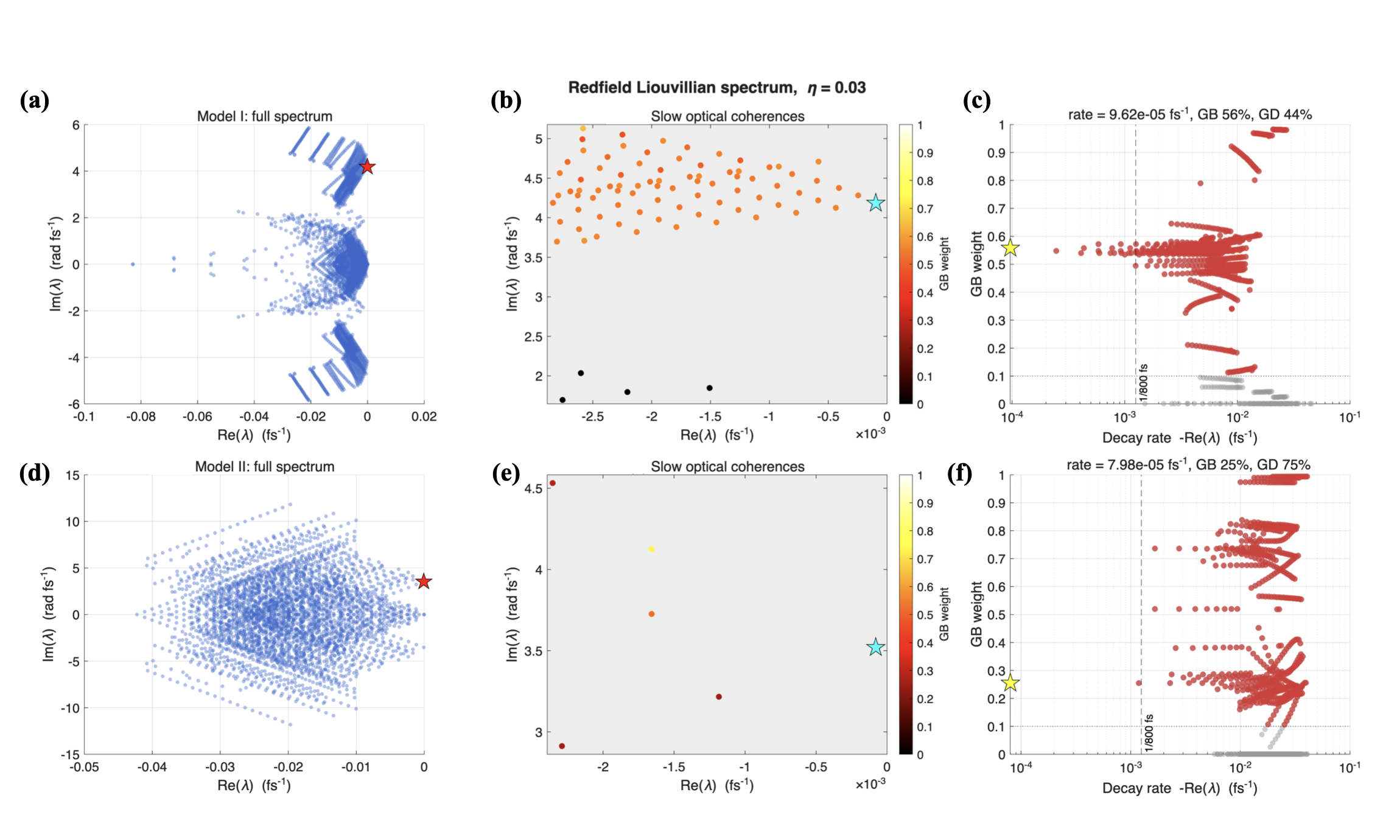}
    \caption{Redfield Liouvillian spectra at $\eta=0.03$ for Model I (top) and Model II (bottom). (a,d) Full spectrum; the star marks the slowest ground-bright coherence mode. (b,e) Slow end of the optical-coherence band, coloured by ground-bright weight $w_{gb}$. (c,f) Ground-bright weight versus decay rate for all optical coherences; red points have $w_{gb}>0.1$, the dashed line marks $1/T$ with $T=800$ fs. The selected mode decays at $9.6\times10^{-5}$ fs$^{-1}$ ($w_{gb}=0.56$, $w_{gd}=0.44$) in Model I and $8.0\times10^{-5}$ fs$^{-1}$ ($w_{gb}=0.25$, $w_{gd}=0.75$) in Model II. Spectra for $\eta=0.10$ and $0.30$ are given in the Supplementary Information (SI).}
    \label{fig:liouvillianspectrum}
\end{figure}

The Liouvillian spectrum is the set of eigenvalues $\lambda_j=-\kappa_j+i\Omega_j$ of the Liouvillian superoperator. The real part gives the decay rate $\kappa_j$ of the corresponding eigenmode, and the imaginary part its oscillation frequency $\Omega_j$. Figure \ref{fig:liouvillianspectrum}(a) and (d) shows the spectra of both models at weak system-bath coupling, $\eta=0.03$, for the same molecular and bath parameters as in the dynamics simulations above, with the real part on the horizontal axis and the imaginary part on the vertical axis. Both spectra contain a dense manifold of rapidly decaying modes and a small number of modes lying much closer to the imaginary axis ($\mathrm{Re}\,\lambda_j\approx0$), which correspond to long-lived components of the density matrix. Modes with $\hbar\Omega_j$ close to the electronic gap are optical coherences; those at the slow end of this band are shown in panels (b) and (e).
\vspace{0.1cm}

A slow mode is relevant for the target observable only if it carries a substantial ground-bright electronic coherence. For each eigenmode $\sigma_j$ we therefore compute its ground-bright weight $w_{gb}=\big(\|\bra{b}\sigma_j\ket{g}\|^2+\|\bra{g}\sigma_j\ket{b}\|^2\big)/\|\sigma_j\|^2$, evaluated in the diabatic basis, and analogously its ground-dark weight $w_{gd}$. Panels (c) and (f) show $w_{gb}$ against the decay rate for all optical coherences. In both models a single ground-bright coherence mode is clearly separated from the rest of the band. We denote its decay rate by $\Gamma$. At $\eta=0.03$, $\Gamma=9.6\times10^{-5}$ fs$^{-1}$ ($1/\Gamma\approx 10$ ps) in Model I and $8.0\times10^{-5}$ fs$^{-1}$ ($1/\Gamma\approx 12.5$ ps) in Model II, about three and fifteen times more slowly, respectively, than the next ground-bright coherence, and two to three orders of magnitude more slowly than the bulk of the band. Only a fraction of the initial coherence is transferred into this mode; the remaining components decay within a few hundred femtoseconds, and it is this fraction that the control maximizes. The mode is not a pure ground-bright coherence. It carries $w_{gb}=0.56$ and $w_{gd}=0.44$ in Model I, and $w_{gb}=0.25$ and $w_{gd}=0.75$ in Model II, consistent with the different degree of mixing each model carries.

\begin{figure}[H]
    \centering
    \includegraphics[width=0.75\textwidth]{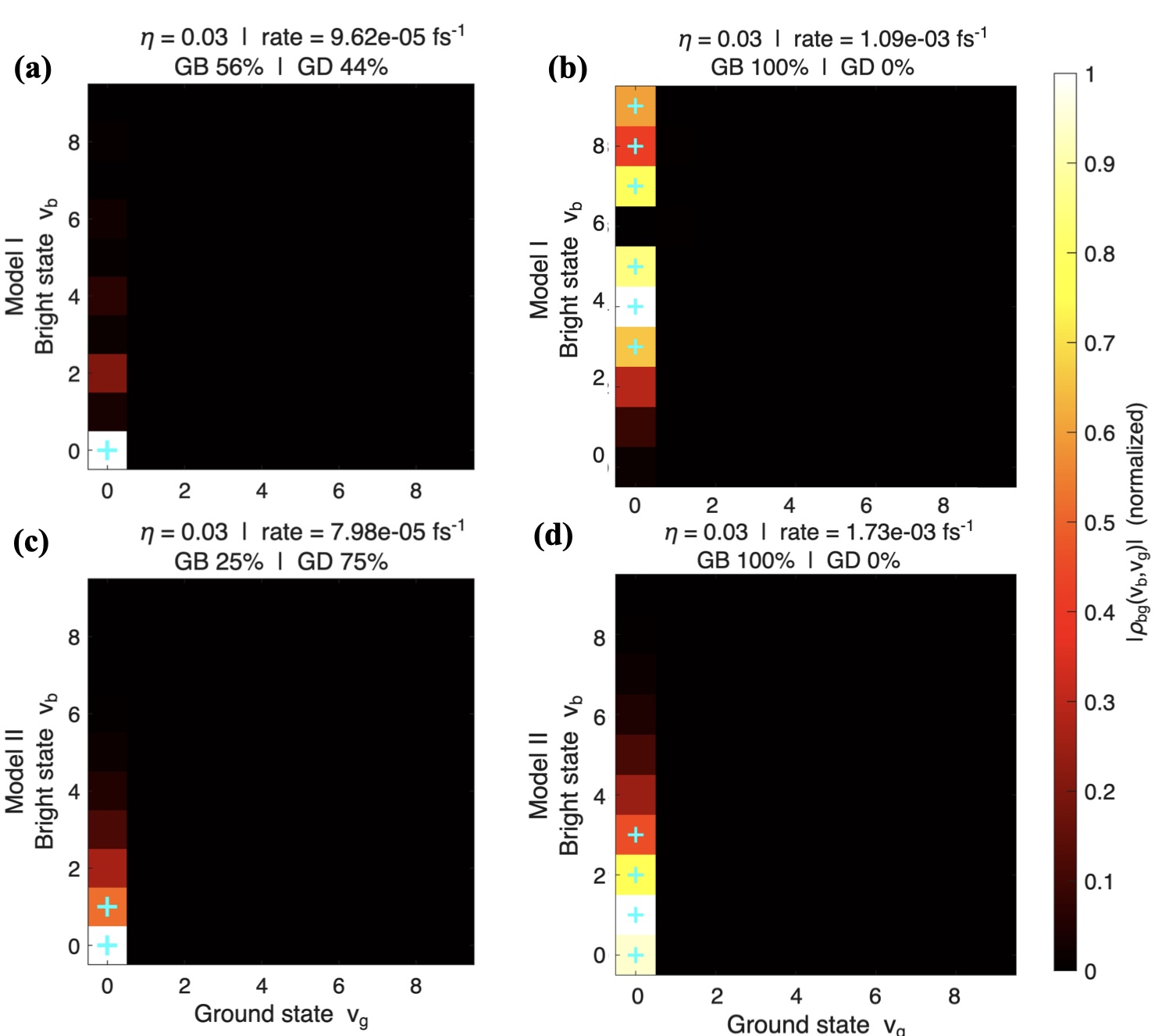}
    \caption{Vibrational structure of the slowest ground-bright coherence mode, $|\rho_{bg}(v_b,v_g)|$ in the diabatic basis (normalized), at $\eta=0.03$.  (a, b) Model I and (c, d) Model II, with (a, c) and without (b, d) the dark-bright coupling. The narrower progression in (c) compared with (d) reflects the smaller excited-state displacement: mixing with the dark state reduces it from $\Delta q=1.50$ to $0.31$ and the decay rate from $1.73\times10^{-3}$ to $7.98\times10^{-5}$ fs$^{-1}$. In (b) the selected mode lies at $3.01$ eV and extends to the edge of the vibrational basis; it is shown for completeness only and is not used in the discussion (SI).}
    \label{fig:slow_modes}
\end{figure}

\vspace{-0.1cm}
Figure \ref{fig:slow_modes} shows the projection of this mode onto the vibrational levels in the Hilbert space. The coherence $|\rho_{bg}(v_b,v_g)|$ is non-zero only for $v_g=0$, so the mode connects the vibrational ground state $|g,v=0\rangle$ with a single excited eigenstate. In the diabatic basis, the bright component of this eigenstate has a maximum at $v_b=0$ and only a short Franck-Condon progression in $v_b$. Its geometry is therefore close to that of the ground state, and since the bath couples to the nuclear coordinate, the two branches of the coherence are hard for the bath to distinguish. The two models reach this situation in different ways. In Model II the excited eigenstate is the lowest state of the excited manifold, at $2.317$ eV, below the origin of the bright state. It is mainly of dark character, and the dark state is almost undisplaced ($s_d=-0.10$), so the eigenstate inherits a ground-like geometry. Being the lowest excited state, it also cannot relax to lower states. In Model I the eigenstate lies higher, at $2.754$ eV, and both diabatic states are strongly displaced, but in opposite directions ($s_d=-2.8$ and $s_b=2.5$). In their mixture the two displacements partly cancel, and the mean geometry is again close to $Q=0$. For $\eta=0.03$, $0.10$ and $0.30$, the decay rate of the mode grows linearly with the coupling strength, with $\Gamma/\eta \approx 3.2\times10^{-3}$ fs$^{-1}$ in Model I and $2.7\times10^{-3}$ fs$^{-1}$ in Model II, while its composition and vibrational structure do not change (SI). The mode is thus a property of the molecular Hamiltonian, not something created by the bath. It is not decoupled from the bath, which would give $\Gamma=0$ at any $\eta$, but only weakly coupled to it, and this is why it decays slowly.
\vspace{0.1cm}

The origin of this weak coupling is geometric. When the excited eigenstate of the coherence is the lowest state of the excited manifold, as in Model II, population relaxation out of it is negligible and the decay is dominated by pure dephasing. Its rate is set by how strongly the bath distinguishes the two branches of the coherence, $\Gamma \simeq \eta\, k_BT\, \Delta q^2$, where $\Delta q$ is the difference in mean vibrational coordinate between the excited and ground eigenstates. This expression follows directly from the diagonal ($m=n$) term of Eq. (\ref{eq:rates_explicit}). For a purely bright excited state $\Delta q = s_b$, whereas the dark state of Model II is almost undisplaced ($s_d=-0.10$). When the two states mix, the dark state lends its geometry to the excited eigenstate and pulls it towards that of the ground state. We verified this directly by switching off the dark-bright coupling ($\Delta_{db}=0$) in Model II. As shown in panel (d), the slowest ground-bright coherence then becomes the bare 0-0 coherence ($w_{gb}=1$, $\hbar\Omega=2.500$ eV) with $\Delta q = 1.50$, and decays at $1.73\times10^{-3}$ fs$^{-1}$, in exact agreement with $\eta k_BT\Delta q^2$. With the dark-bright coupling restored, $\Delta q$ drops to $0.31$ and the rate to $7.98\times10^{-5}$ fs$^{-1}$, consistent with the $\Delta q^2$ scaling, $(1.50/0.31)^2\approx 23$. The same effect is visible in the vibrational structure of the mode. Without the dark state the Franck-Condon progression extends from $v_b=0$ to $v_b=4$, as expected for a displacement of $1.5$ (Fig. \ref{fig:slow_modes}(d)), whereas the protected mode is confined essentially to $v_b\le 2$ (Fig. \ref{fig:slow_modes}(c)). The dark state thus acts as a shield against decoherence: it gives the optically accessible excited state the geometry of the ground state, so that the bath can hardly tell the two branches of the coherence apart. The mere presence of a dark state is however not sufficient. Protection requires a dark state that is weakly displaced with respect to the ground state and strongly mixed with the bright state. These two conditions can be stated in terms of quantities that are directly evaluated with electronic-structure calculations, such as excited-state displacements and diabatic couplings, and they provide a criterion to search for real molecules in which such protected coherences exist. In Model I the displacement of the protected state is even smaller ($\Delta q=0.16$), and its decay is instead dominated by bath-induced thermal transitions out of $|g,v=0\rangle$, which are significant because the vibrational quantum ($0.04$ eV) is comparable to $k_BT$ (SI).
\vspace{0.1cm}

Most vibronic modes decay rapidly, and an uncontrolled excitation spreads the coherence over many of them. The available resource is thus not the absence of dissipation, but a single, well-isolated, dark-state-protected mode that survives for picoseconds. Whether it is used depends on how the system is excited, which we examine next.

\subsection{The Liouvillian spectrum predicts how to access the protected coherence}
\label{subsec:dissipation_results}

\begin{figure}[H]
    \centering
    \includegraphics[width=0.8\textwidth]{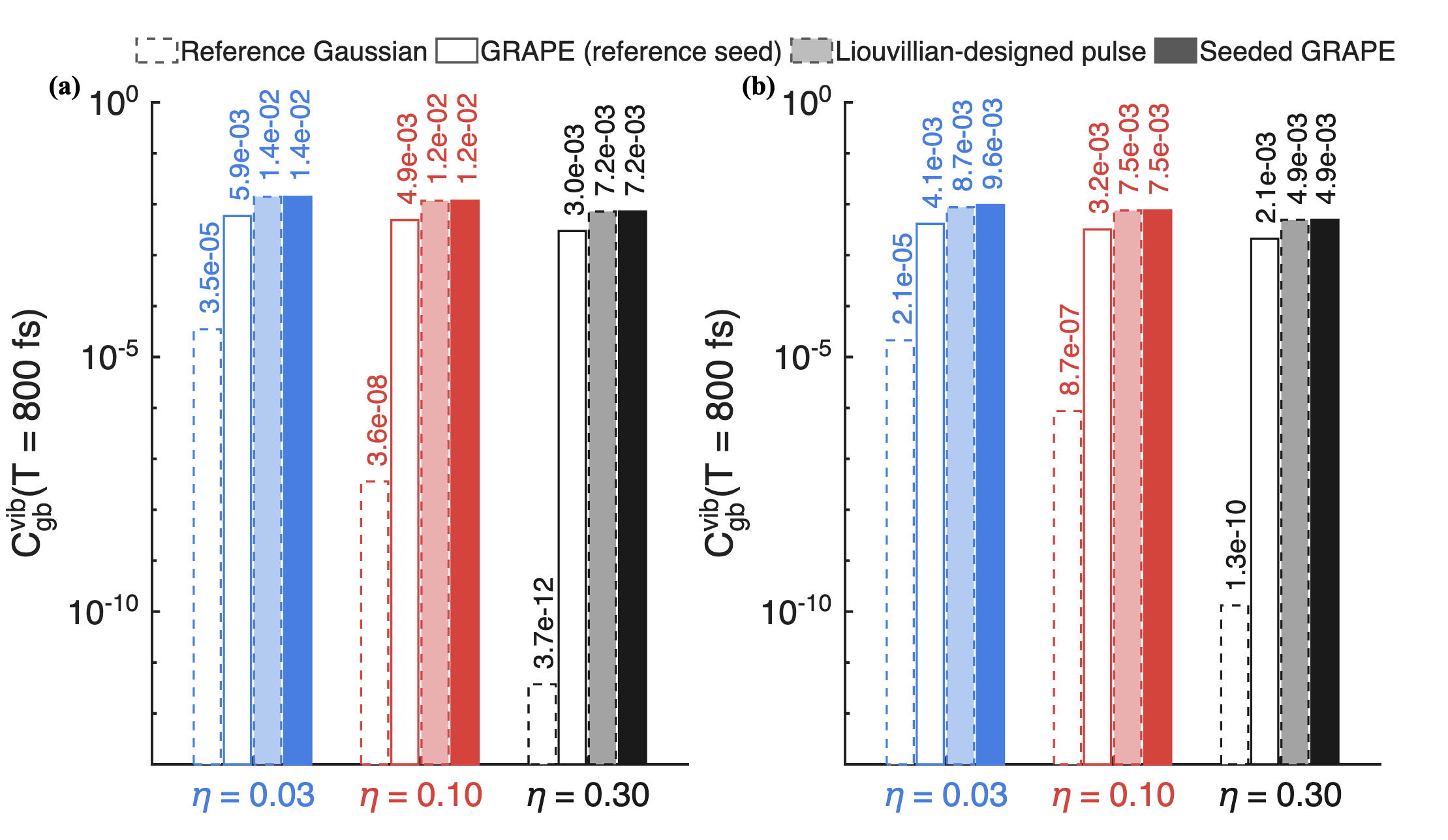}
    \caption{Ground-bright coherence $\Cgb^{\mathrm{vib}}(T)$ at $T=800$ fs for the reference Gaussian pulse ($2.53$ eV), the GRAPE-optimized pulse, the Liouvillian-designed pulse (Gaussian tuned to the protected mode, $2.754$ eV in Model I and $2.317$ eV in Model II), and GRAPE initialized with the Liouvillian-designed pulse (seeded GRAPE), all of equal fluence, at $\eta=0.03$ (blue), $0.10$ (red) and $0.30$ (black). Dashed edges: Gaussian pulses; solid edges: GRAPE. Empty bars: reference seed; filled bars: protected-mode seed. (a) Model I. (b) Model II. Logarithmic vertical axis.}
    \label{fig:gain}
\end{figure}

The protected mode identified in Sec. \ref{subsec:liouvillian_spectrum} has a well-defined optical frequency, $\hbar\Omega = 2.754$ eV in Model I and $2.317$ eV in Model II. Both values correspond to optical transitions from the vibrational ground state: the protected state carries 53\% (Model I) and 19\% (Model II) of the oscillator strength from $|g,v=0\rangle$, and in Model I it is the most intense transition (SI). We therefore compare four excitation protocols of identical fluence: (i) the Gaussian reference pulse, centred at $2.53$ eV, slightly energy-detuned above the minima of the bright state; (ii) the same Gaussian pulse tuned to the frequency of the protected mode, which we refer to as the Liouvillian-designed pulse; (iii) a GRAPE-optimized pulse obtained starting from the reference pulse (i), without any information about the Liouvillian spectrum; and (iv) a GRAPE-optimized pulse seeded with the Liouvillian-designed pulse (ii) (seeded GRAPE, Sec. \ref{subsec:grape}). The coherence $\Cgb^{\mathrm{vib}}(T)$ is evaluated at the target time of the optimization, $T=800$ fs, for $\eta=0.03$, $0.10$ and $0.30$ (Table \ref{tab:protocols} and Fig. \ref{fig:gain}).

\begin{table}[H]
\centering

\caption{Ground-bright coherence $\Cgb^{\mathrm{vib}}(T)$ at $T=800$ fs for four excitation protocols of equal fluence. The Liouvillian-designed pulse is the reference Gaussian tuned to the protected mode; seeded GRAPE is initialized with it.}
\label{tab:protocols}
\vspace{0.3cm}

\begin{tabular}{llcccc}
\hline\hline
 & $\eta$ & Reference Gaussian & GRAPE & Liouvillian-designed pulse & Seeded GRAPE \\
\hline
Model I  & 0.03 & $3.5\times10^{-5}$  & $5.9\times10^{-3}$ & $1.41\times10^{-2}$ & $1.41\times10^{-2}$ \\
         & 0.10 & $3.6\times10^{-8}$  & $4.9\times10^{-3}$ & $1.17\times10^{-2}$ & $1.17\times10^{-2}$ \\
         & 0.30 & $3.7\times10^{-12}$ & $3.0\times10^{-3}$ & $7.21\times10^{-3}$ & $7.21\times10^{-3}$ \\
\hline
Model II & 0.03 & $2.1\times10^{-5}$  & $4.1\times10^{-3}$ & $8.74\times10^{-3}$ & $9.63\times10^{-3}$ \\
         & 0.10 & $8.7\times10^{-7}$  & $3.2\times10^{-3}$ & $7.54\times10^{-3}$ & $7.54\times10^{-3}$ \\
         & 0.30 & $1.3\times10^{-10}$ & $2.1\times10^{-3}$ & $4.94\times10^{-3}$ & $4.94\times10^{-3}$ \\
\hline\hline
\end{tabular}
\end{table}

The reference pulse is not resonant with the protected mode and therefore populates mainly fast-decaying optical coherences. Its coherence is already small at the target time ($\sim\!10^{-5}$ for $\eta=0.03$, empty dashed bars in Fig. \ref{fig:gain}) and decreases by several orders of magnitude more at stronger coupling.

Tuning the same pulse to the protected mode, which we refer to as the Liouvillian-designed pulse, changes the picture completely. The resulting coherence exceeds that of the reference pulse by a factor of about $400$ at $\eta=0.03$ in both models, and by much more at stronger coupling (Fig. \ref{fig:gain} and Table \ref{tab:protocols}). If this long-lived coherence is carried by the protected mode, its value at $T$ is $\Cgb^{\mathrm{vib}}(T;\eta)\propto e^{-\Gamma_\eta T}$, with a prefactor set by the pulse and independent of $\eta$, where $\Gamma_\eta$ is the decay rate of the protected mode obtained from the Liouvillian spectrum (Sec. \ref{subsec:liouvillian_spectrum}). The ratios $\Cgb^{\mathrm{vib}}(T;\eta)/\Cgb^{\mathrm{vib}}(T;0.03)$ should therefore equal $\exp[-(\Gamma_\eta-\Gamma_{0.03})\,T]$. The simulated ratios are $0.835$ and $0.513$ in Model I (predicted $0.836$ and $0.510$), and $0.862$ and $0.565$ in Model II (predicted $0.862$ and $0.566$), without any adjustable parameter; the agreement is better than 1\% in all cases. The coherence also decays somewhat faster than $\Gamma$ alone, by an amount that is independent of $\eta$: about $3.6\times10^{-4}$ fs$^{-1}$ in Model I and $1.3\times10^{-4}$ fs$^{-1}$ in Model II. We attribute this additional loss to the finite bandwidth of the pulse, which also excites neighbouring optical coherences that oscillate at slightly different frequencies and dephase relative to the protected mode. Consistent with this interpretation, the loss is smaller in Model II, where the protected mode is spectrally better isolated (the next ground-bright coherence decays about fifteen times faster, compared with about three times in Model I). This observation confirms that the bath-induced decay of the long-lived coherence is that of the single dark-state-protected eigenmode, and that this mode is directly accessible by resonant optical excitation.
\vspace{0.1cm}

The GRAPE-optimized pulse starts from the detuned reference and has no knowledge of the Liouvillian spectrum. It nevertheless finds the protected mode on its own: the optimized coherence decays at essentially the same rate as that produced by the tuned pulse, and exceeds the reference by more than two orders of magnitude at $\eta=0.03$ and by many more at stronger coupling, where the reference coherence has vanished. In Model I its ratios $\Cgb^{\mathrm{vib}}(T;\eta)/\Cgb^{\mathrm{vib}}(T;0.03)$ are $0.832$ and $0.508$, again within 1\% of the prediction. Optimal control therefore discovers the long-lived, dark-state-protected coherence by its own. Its yield, however, remains 2.1 to 2.4 times below that of the pulse tuned to the protected mode, nearly independently of the model and of the coupling strength. This is a consequence of the reference optimization protocol, not a limitation of the control. Two features of this protocol are responsible. First, the fluence is not kept fixed during the optimization (iii): the optimized pulse reaches a fluence about two orders of magnitude larger than the reference pulse and is rescaled to the reference fluence only at the end. Second, the Blackman window is applied to the full pulse at every iteration, which progressively shortens the pulse and broadens its spectrum. Because the coherence at $T$ is linear in the field amplitude, the rescaled yield measures the quality of the pulse shape alone. This yield is largest after about ten iterations ($8.1\times10^{-3}$ in Model I at $\eta=0.03$) and decreases afterwards, while the objective $J$, which includes the fluence penalty, keeps converging. The last iterate, which is the one reported, is therefore not the best pulse shape found during the optimization.
\vspace{0.1cm}

The Liouvillian spectrum thus provides what is needed to address the protected coherence: its existence, frequency and lifetime. With this information, a transform-limited pulse tuned to the protected mode populates it efficiently, and the resulting coherence survives for picoseconds even at strong dissipation. Initializing GRAPE with this Liouvillian-designed pulse at fixed fluence (Sec. \ref{subsec:grape}) increases the coherence at $T$ by 10\% in Model II at $\eta=0.03$, from $8.74\times10^{-3}$ to $9.63\times10^{-3}$, and by less than 0.1\% in all other cases (Table \ref{tab:protocols} and Fig. \ref{fig:gain}). Since the seeded optimization cannot fall below its initial pulse, the designed pulse is already close to the best pulse within the reach of the gradient optimization, and further shaping adds little once the protected mode is addressed.


\section{Discussion}
\label{sec:discussion}

The results show that optimal control can directly target electronic coherence in an dissipative vibronic system by orders of magnitude at a fixed fluence, and they identify where this enhancement comes from. The field does not avoid or suppress the bath. The molecular system contains a coherence mode that is only weakly coupled to the bath, and the role of the field is to transfer as much of the initial coherence as possible into this mode.
\vspace{0.1cm}

This changes how the control problem should be read. The enhancement factors in Table \ref{tab:protocols} are large mainly because the reference pulse, centred near the bright-state origin, populates fast-decaying coherences: at strong coupling its coherence at $T$ has essentially vanished, so that any pulse that reaches the protected mode wins by many orders of magnitude. The more informative comparison is with the pulse tuned to the protected mode, which shows that a simple transform-limited pulse already captures most of what is available once the right frequency is known. The role of the Liouvillian spectrum is precisely to provide this information: which coherences survives, at which frequencies they can be excited, and how their lifetime scales with the coupling to the bath. GRAPE, without this information, arrives at the same mode on its own, so the spectral analysis point to the same physics from two independent directions. Combining the two, the Liouvillian to locate the protected mode and optimal control at fixed fluence to refine the pulse, performs best, but most of the gain comes from knowing where the protected coherence lives.
\vspace{0.1cm}

Several features distinguish this problem from more conventional applications of quantum optimal control. First, the target is directly a coherence, not a population, so the optimizer must control both the amplitude and the phase of a coupled electronic-nuclear wave packet rather than simply steering probability from one state to another. Second, the control is performed directly against dissipation: the Redfield contribution enters both the forward propagation and its adjoint, so the resulting gradient already encodes how the field influences coherent and dissipative evolution together. An optimization performed against the isolated molecular Hamiltonian could produce a field that performs well in the absence of the bath but loses its advantage once dissipation is reintroduced. Third, the nuclear degree of freedom is not only the channel through which decoherence occurs, but also the reason why some coherences are protected: since the bath couples to the nuclear coordinate, the geometry of the two states involved, and therefore the dependence with $Q$, determines how fast their coherence decays, and the optimized field keeps the excited-state wave packet close to the ground-state geometry (Fig.
\ref{fig:wavepacket}).
\vspace{0.1cm}

The protection mechanism itself is simple and, in our view, its most useful aspect. Mixing with a weakly displaced dark state pulls the geometry of an optically accessible excited state towards that of the ground state, so that the bath can hardly distinguish the two branches of the coherence. This gives a molecular design criterion that can be stated in terms of system parameters: a dark state that is weakly displaced with respect to the ground state and strongly mixed with the bright state. Usually, dark states are regarded as a loss channel for optically prepared excitations, although vibronic mixing has been invoked to explain long-lived electronic coherence in light-harvesting systems \cite{lim_vibronic_2015}. Here they are what makes a long-lived optical coherence possible. On the basis of the present work, we propose vibronic molecules, with their chemically tunable dark-state structure, as natural carriers of long-lived coherence for quantum information applications.
\vspace{0.1cm}

The present work nonetheless has some limits that need to be stated explicitly. Redfield theory is a second-order, weak-coupling description of the system-bath interaction, and its accuracy is expected to degrade for systems with strongly non-Markovian, or highly structured environments \cite{Jones-Nonmarkovian}. The pure-dephasing estimate $\Gamma\simeq\eta k_BT\Delta q^2$ is likewise a Markovian, high-temperature result. The present results therefore should be read as establishing the mechanism within the Markovian limit, rather than as a general statement about molecular decoherence at arbitrary coupling strength. Our optimal-control framework is not tied to Redfield theory, however. The dissipative propagator and its adjoint could be replaced by a nonperturbative formalism, such as hierarchical equations of motion \cite{tanimura_numerically_2020}, to test whether the dark-state protection persists outside the perturbative regime. We are currently pursuing this extension.
\vspace{0.1cm}

A second outlook is the investigation beyond a single explicitly treated vibrational coordinate. Real molecules have many modes, and the bath may couple to several of them. A protected coherence then requires the dark-state mixing to reduce the displacement along all strongly coupled modes, which is a natural question for multidimensional models. The present work establishes the physical principles in a minimal vibronic model and lays the foundation for extensions to more complex molecular systems. Finally, any experimental realization will need to respect finite laser bandwidth, pulse-shaper resolution, maximum intensity and spectral filtering. Some of these properties are contained in our study within the Blackman Window implementation. Further constraints can be, however, considered directly into the objective functional we introduce. In this respect, the additional observation that a transform-limited pulse at the right frequency can reach the protected mode is also encouraging for experiments.

\section{Conclusions}
\label{sec:conclusions}

This work asks whether electronic coherence in dissipative vibronic systems is a fixed casualty of the environment, or whether part of it can be preserved by the way the molecule is excited. To answer this, a microscopic vibronic Hamiltonian, Redfield dissipative dynamics and gradient-based GRAPE optimization are combined, with the photoinduced vibronic coherence at a deliberately late target time of $800$ fs as the control objective, and the spectrum of the molecular Liouvillian as the tool to interpret the result.
\vspace{0.1cm}

In two vibronic models chosen to be dynamically different, the control fields generate a much larger long-time coherence than a conventional Gaussian excitation of the same fluence, at all coupling strengths considered. Both models contain a single coherence mode, of mixed ground-bright and ground-dark character, that decays on a picosecond timescale because mixing with the dark state brings the geometry of the excited state close to that of the ground state. The optimized field finds this mode without being told in which are of the vibronic space is, keeping the excited-state wave packet near the Franck-Condon region, and the Liouvillian predicts how the resulting coherence depends on the coupling to the bath to within 1\%. A Gaussian pulse tuned to this mode reaches it almost optimally.
\vspace{0.1cm}

The broader implication is that the dissipative spectrum of a molecule can be treated as a map for control. Here we suggest that instead of asking how fast coherence decays, one can ask which coherences decay slowly, why, and at which frequency they can be addressed. In vibronic molecules, the answer is tied to the dark states and to the geometry of the excited-state manifold, which can in principle be tuned chemically \cite{hayes_engineering_2013}. This makes vibronic dark states a built-in resource for long-lived electronic coherence. The approach we introduce here is therefore general: it requires only the Liouvillian of the system and a coherence control objective, applying to any vibronic system and any bath model for which these can be constructed.
\vspace{0.1cm}

These results establish molecular vibronic systems as a potential resource for long-lived quantum coherence and open the way to encoding quantum information in protected vibronic subspaces. The next steps are to identify molecules that meet the design criterion reported in this work, to extend the analysis to multidimensional systems and nonperturbative bath-system conditions.

\begin{acknowledgments}
The author thanks Prof. Ignacio R. Sola and Dr. Garth Jones for stimulating discussions on optimal control theory and dissipative dynamics, and Dr. David Picconi for his support.
\end{acknowledgments}

\bibliography{references_paper}

\end{document}


\maketitle
\tableofcontents

\section{Liouvillian spectra at stronger system-bath coupling}
\label{si:spectra}

\begin{figure}[H]
\centering
\includegraphics[width=\textwidth]{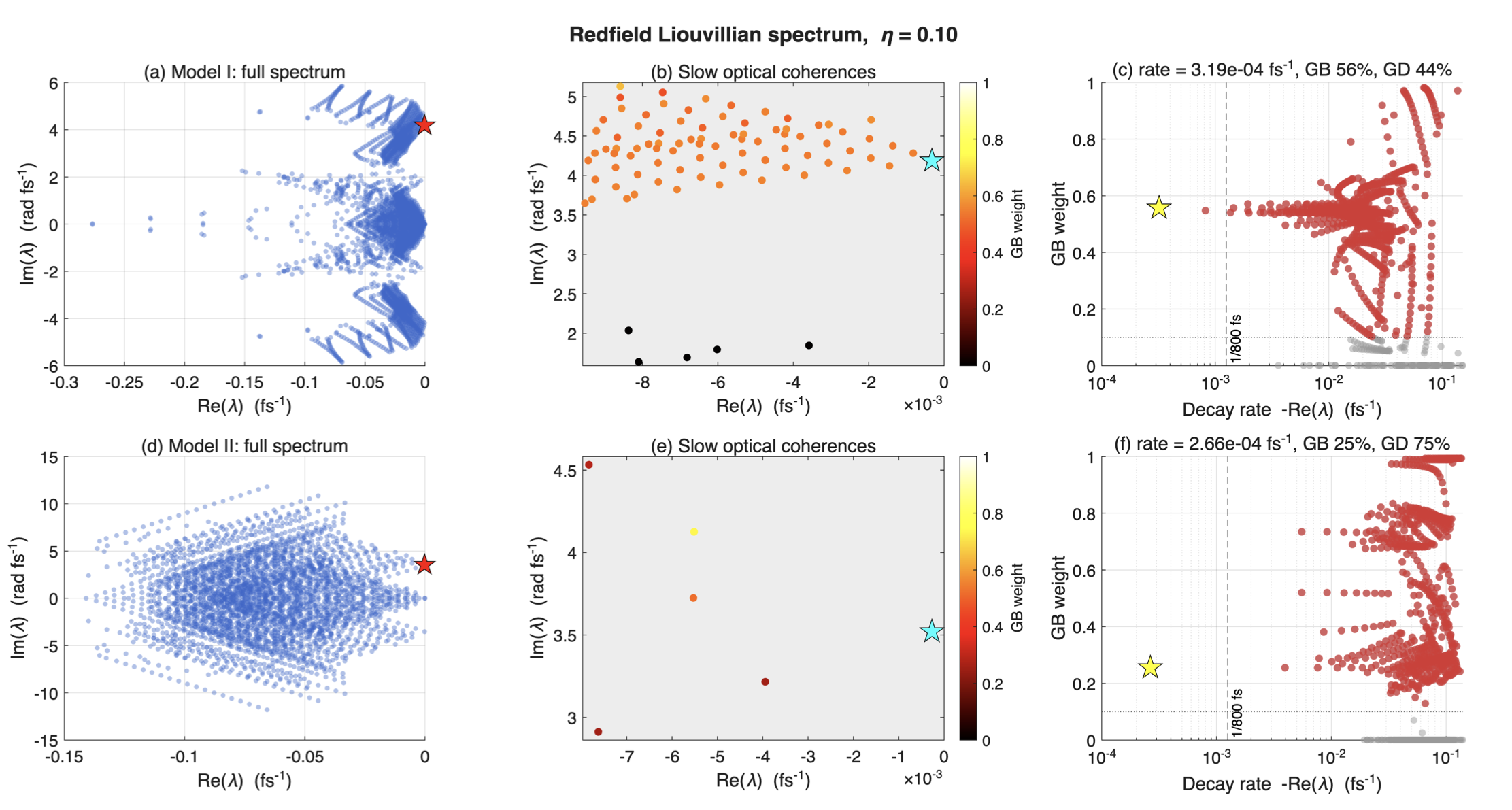}
\caption{Redfield Liouvillian spectra at $\eta=0.10$ for Model I (top) and Model II (bottom). Panels as in Fig. 5 of the main text.}
\label{fig:si_spectrum_010}
\end{figure}

\begin{figure}[H]
\centering
\includegraphics[width=\textwidth]{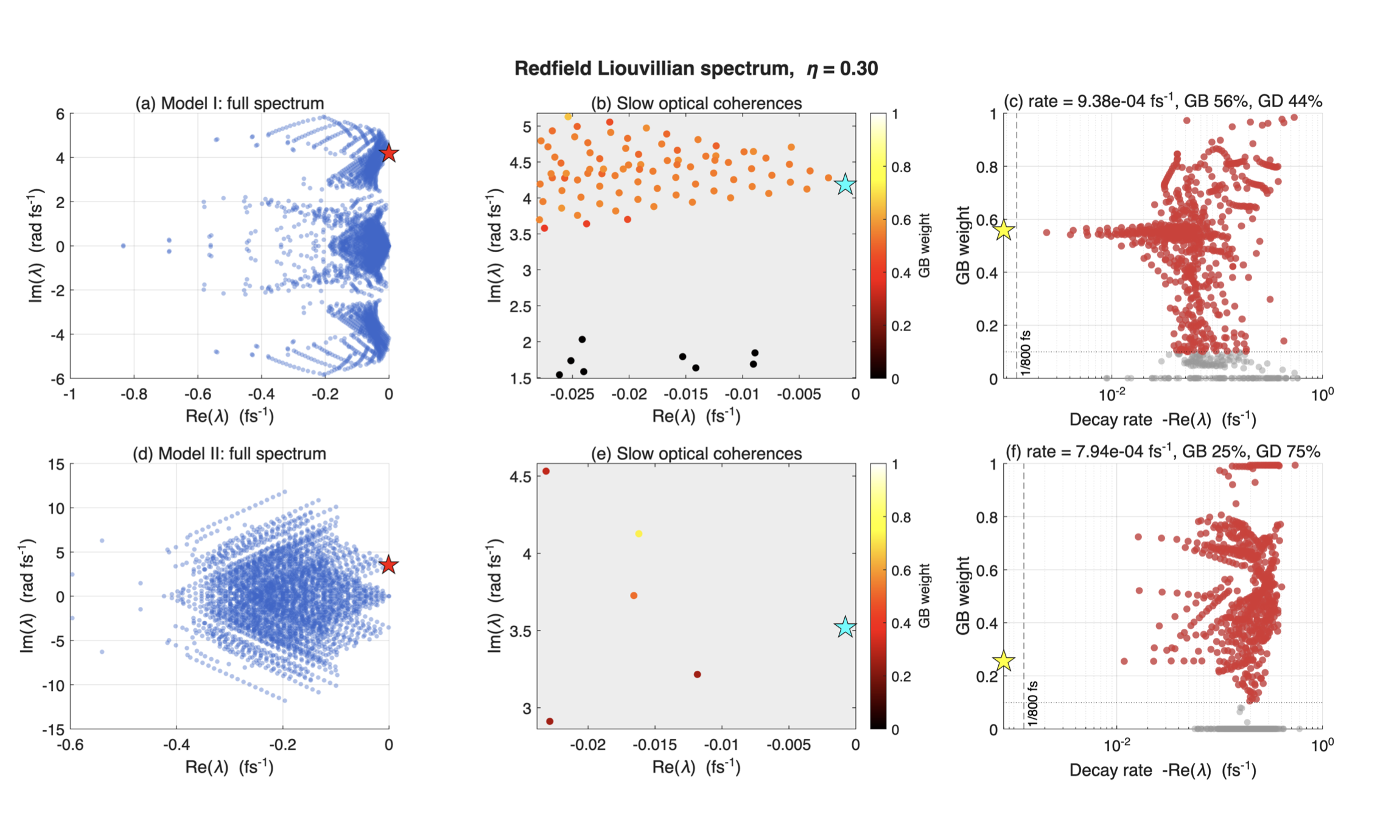}
\caption{Redfield Liouvillian spectra at $\eta=0.30$ for Model I (top) and Model II (bottom). Panels as in Fig. 5 of the main text.}
\label{fig:si_spectrum_030}
\end{figure}

Figures \ref{fig:si_spectrum_010} and \ref{fig:si_spectrum_030} show the Redfield Liouvillian spectra for $\eta=0.10$ and $\eta=0.30$, in the same format as Fig. 5 of the main text ($\eta=0.03$). In both models the same ground-bright coherence is selected as the slowest optical coherence with $w_{gb}>0.1$. Its oscillation frequency and its composition are unchanged, and only its decay rate grows with $\eta$ (Table \ref{tab:si_modes}).

\begin{table}[H]
\centering
\caption{Slowest ground-bright optical coherence of the Redfield Liouvillian for the three coupling strengths.}
\label{tab:si_modes}
\vspace{0.3cm}

\begin{tabular}{llccccc}
\toprule
 & $\eta$ & $\Gamma$ (fs$^{-1}$) & $\Gamma/\eta$ (fs$^{-1}$) & $1/\Gamma$ (ps) & $\hbar\Omega$ (eV) & $w_{gb}$ / $w_{gd}$ \\
\midrule
Model I  & 0.03 & $9.6\times10^{-5}$  & $3.21\times10^{-3}$ & 10.4 & 2.754 & 0.56 / 0.44 \\
         & 0.10 & $3.2\times10^{-4}$  & $3.20\times10^{-3}$ & 3.1  & 2.754 & 0.56 / 0.44 \\
         & 0.30 & $9.4\times10^{-4}$  & $3.13\times10^{-3}$ & 1.1  & 2.754 & 0.56 / 0.44 \\
\midrule
Model II & 0.03 & $8.0\times10^{-5}$  & $2.66\times10^{-3}$ & 12.5 & 2.317 & 0.25 / 0.75 \\
         & 0.10 & $2.66\times10^{-4}$ & $2.66\times10^{-3}$ & 3.8  & 2.317 & 0.25 / 0.75 \\
         & 0.30 & $7.9\times10^{-4}$  & $2.65\times10^{-3}$ & 1.3  & 2.317 & 0.25 / 0.75 \\
\bottomrule
\end{tabular}
\end{table}

\section{Vibrational structure of the protected mode for all coupling strengths}
\label{si:maps}

\begin{figure}[H]
\centering
\includegraphics[width=\textwidth]{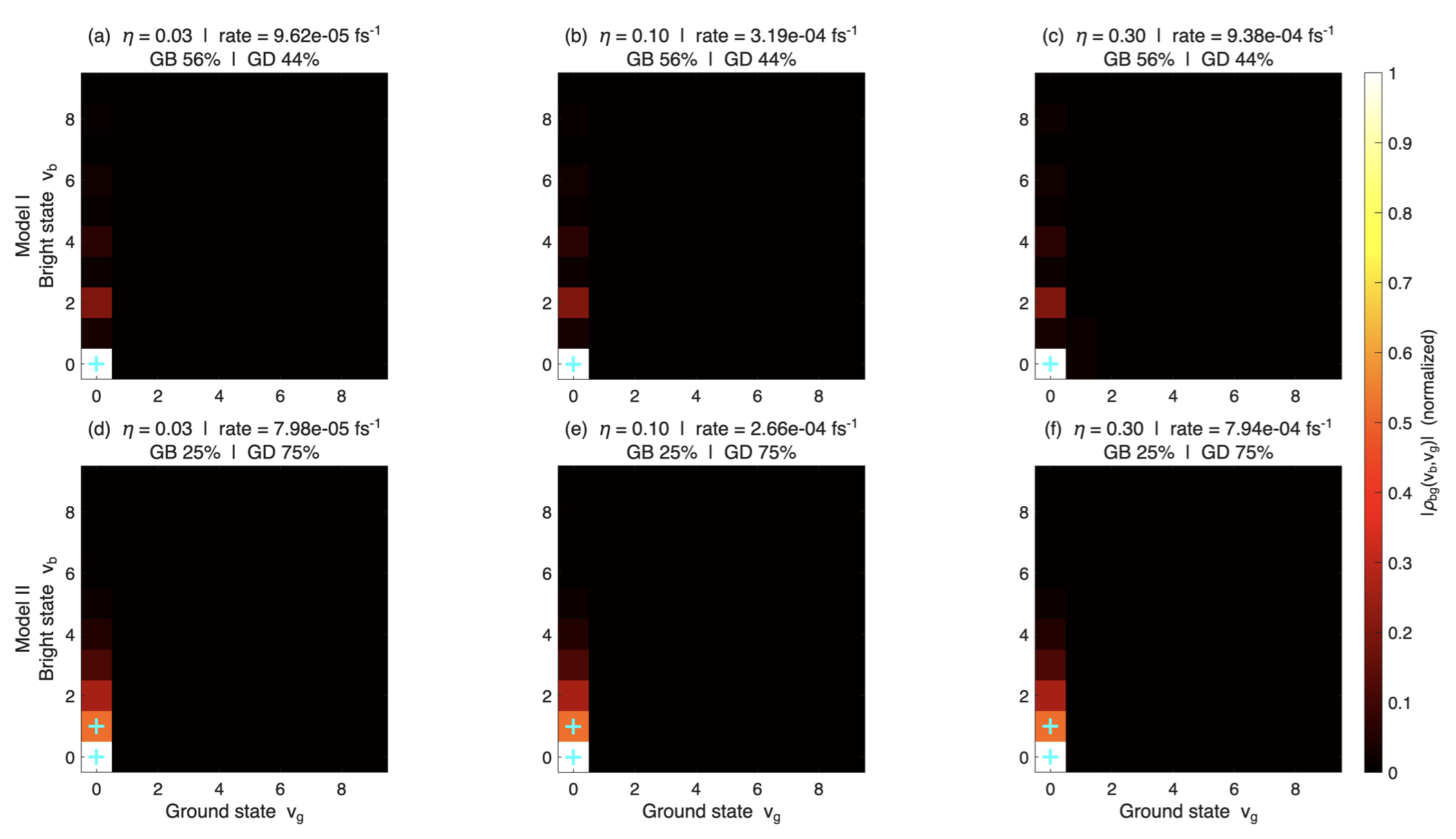}
\caption{Normalized $|\rho_{bg}(v_b,v_g)|$ of the slowest ground-bright coherence mode for $\eta=0.03$, $0.10$ and $0.30$ (columns), for Model I (top) and Model II (bottom). Crosses mark cells with normalized amplitude above $0.3$.}
\label{fig:si_maps}
\end{figure}

Figure \ref{fig:si_maps} shows $|\rho_{bg}(v_b,v_g)|$ of the protected mode, in the diabatic basis, for the three coupling strengths. The vibrational structure is the same for all $\eta$: the mode is confined to $v_g=0$ and peaks at $v_b=0$. The system-bath coupling thus sets the decay rate of the mode but not its composition. This justifies taking the prefactor of the long-time coherence as independent of $\eta$ in the prediction of the main text.

\section{Convergence with respect to the vibrational basis}
\label{si:convergence}

All calculations use 20 vibrational levels per electronic state. To verify that the protected mode is converged in this basis, we evaluate the fraction of its ground-bright component that lies in the four highest vibrational levels,
\begin{equation}
f_{\mathrm{edge}} = \frac{\sum_{v_b\ge16\ \mathrm{or}\ v_g\ge16} |\rho_{bg}(v_b,v_g)|^2}{\sum_{v_b,v_g} |\rho_{bg}(v_b,v_g)|^2}.
\end{equation}
We obtain $f_{\mathrm{edge}} = 2.0\times10^{-4}$, $1.4\times10^{-4}$ and $5.3\times10^{-5}$ for Model I and $f_{\mathrm{edge}}\sim10^{-15}$ for Model II, at $\eta=0.03$, $0.10$ and $0.30$. The protected mode does not reach the edge of the basis, so its rate and structure are converged.

\section{Role of the dark state: calculation without dark-bright coupling}
\label{si:nodark}

To isolate the contribution of the dark state we repeated the analysis of Model II with the dark-bright coupling switched off ($\Delta_{db}=0$), keeping all other parameters fixed. The slowest ground-bright optical coherence is then the bare $0$-$0$ coherence between $|g,v=0\rangle$ and the lowest vibronic level of the bright state, with $w_{gb}=1$, $\hbar\Omega=2.5$ eV and $\Delta q=s_b=1.50$. Its decay rates are $1.73\times10^{-3}$, $5.78\times10^{-3}$ and $1.73\times10^{-2}$ fs$^{-1}$ for $\eta=0.03$, $0.10$ and $0.30$, i.e.\ $\Gamma/\eta=5.78\times10^{-2}$ fs$^{-1}$ for all three. This is exactly the pure-dephasing value $k_BT\,\Delta q^2$, with the prefactor that follows from the spectral density of the main text, that follows from the expression $\Gamma\simeq\eta k_BT\,\Delta q^2$. With $\Delta_{db}=0.15$ eV the rate drops to $8.0\times10^{-5}$ fs$^{-1}$ at $\eta=0.03$, a twenty-two-fold reduction (Table \ref{tab:si_nodark}).

\hspace{0.3cm}

\begin{table}[H]
\centering
\caption{Protected coherence of Model II with and without dark-bright coupling ($\eta=0.03$).}
\label{tab:si_nodark}
\vspace{0.3cm}

\begin{tabular}{lccccc}
\toprule
 & $\hbar\Omega$ (eV) & $w_{gb}$ / $w_{gd}$ & $\Delta q$ & $\eta k_BT\Delta q^2$ (fs$^{-1}$) & $\Gamma$ (fs$^{-1}$) \\
\midrule
$\Delta_{db}=0$         & 2.500 & 1.00 / 0.00 & 1.50 & $1.73\times10^{-3}$ & $1.73\times10^{-3}$ \\
$\Delta_{db}=0.15$ eV   & 2.317 & 0.25 / 0.75 & 0.31 & $7.3\times10^{-5}$  & $8.0\times10^{-5}$  \\
\bottomrule
\end{tabular}
\end{table}

For Model I without dark-bright coupling, the slowest ground-bright coherence selected by the criterion lies at $3.01$ eV and its bright component extends to the edge of the vibrational basis (Fig. 6(b) of the main text). Its rate is therefore not converged, and Model I is not used for this comparison. The $0$-$0$ coherence of Model I without coupling has $\Delta q=s_b=2.5$, corresponding to a pure-dephasing rate $\eta k_BT\Delta q^2 = 4.8\times10^{-3}$ fs$^{-1}$ at $\eta=0.03$, fifty times faster than the protected mode. 

\section{Optical accessibility of the protected mode}
\label{si:oscillator}

The relative strength of the optical transition from $|g,v=0\rangle$ to each excited eigenstate $|k\rangle$ of $H_S$, in the Condon approximation and with the transition dipole coupling the ground state to the bright state only, is
\begin{equation}
f_k = \big|\langle k|\,\hat\mu\,|g,v=0\rangle\big|^2 \Big/ \sum_{k'}\big|\langle k'|\,\hat\mu\,|g,v=0\rangle\big|^2 .
\end{equation}
The most intense transitions are listed in Table \ref{tab:si_osc}. In Model I the excited eigenstate of the protected coherence carries $53\%$ of the oscillator strength and is the most intense transition. In Model II it carries $19\%$ and is the second most intense. The protected mode is accessible by optical excitation in both models investigated.

\begin{table}[H]
\centering
\caption{Most intense transitions from $|g,v=0\rangle$: transition energy, relative oscillator strength $f_k$, and bright-state weight of the excited eigenstate. The excited state of the protected coherence is marked with an asterisk.}

\label{tab:si_osc}
\vspace{0.3cm}

\begin{tabular}{ccc@{\hspace{1.5cm}}ccc}
\toprule
\multicolumn{3}{c}{Model I} & \multicolumn{3}{c}{Model II} \\
$E$ (eV) & $f_k$ & bright weight & $E$ (eV) & $f_k$ & bright weight \\
\midrule
2.754$^*$ & 0.53 & 0.56 & 2.715     & 0.43 & 0.74 \\
2.422     & 0.18 & 0.21 & 2.317$^*$ & 0.19 & 0.25 \\
2.441     & 0.11 & 0.58 & 2.983     & 0.11 & 0.29 \\
2.390     & 0.05 & 0.09 & 2.914     & 0.10 & 0.69 \\
2.818     & 0.04 & 0.55 & 2.770     & 0.05 & 0.27 \\
\bottomrule
\end{tabular}
\end{table}

\section{Decay channels of the protected mode}
\label{si:decay}

The decay rate of the protected mode can be separated into three contributions of the Redfield tensor: pure dephasing, which comes from the diagonal ($m=n$) term and equals $\eta k_BT\,\Delta q^2$; bath-induced transitions out of the excited eigenstate $|e\rangle$ of the coherence; and bath-induced transitions out of $|g,v=0\rangle$. Table \ref{tab:si_decay} gives these contributions at $\eta=0.03$. Their sum reproduces the Liouvillian eigenvalue in both models.

\begin{table}[H]
\centering
\caption{Contributions to the decay rate of the protected mode at $\eta=0.03$ (fs$^{-1}$).}
\label{tab:si_decay}
\vspace{0.3cm}

\begin{tabular}{lcccc}
\toprule
 & pure dephasing & transitions out of $|e\rangle$ & transitions out of $|g,v=0\rangle$ & total \\
\midrule
Model I  & $2.0\times10^{-5}$ & $1.8\times10^{-5}$ & $5.9\times10^{-5}$ & $9.6\times10^{-5}$ \\
Model II & $7.3\times10^{-5}$ & $0.6\times10^{-5}$ & $0.05\times10^{-5}$ & $8.0\times10^{-5}$ \\
\bottomrule
\end{tabular}
\end{table}

In Model II the decay is dominated by pure dephasing, so that the geometric argument of the main text applies directly. In Model I the displacement of the protected state is smaller ($\Delta q=0.16$), pure dephasing is weak, and the decay is dominated by thermal excitation out of $|g,v=0\rangle$, which is significant because the vibrational quantum ($0.04$ eV) is only $1.6\,k_BT$. In Model II ($0.20$ eV, about $8\,k_BT$) this channel is negligible.